\documentstyle[amsfonts,aps,prl,epsf]{revtex}
\newcommand{\beq}{\begin{equation}}
\newcommand{\eeq}{\end{equation}}
\newcommand{\beqa}{\begin{eqnarray}}
\newcommand{\eeqa}{\end{eqnarray}}
\newcommand{\non}{\nonumber}

\begin{document}

\title{Ohta-Jasnow-Kawasaki Approximation for Nonconserved Coarsening 
under Shear}

\author{
Andrea Cavagna\thanks{E-mail:andrea@a13.ph.man.ac.uk},
Alan J. Bray\thanks{E-mail:bray@a13.ph.man.ac.uk} 
and
Rui D. M. Travasso\thanks{E-mail:rui@a13.ph.man.ac.uk} 
}

\address{
Department of Physics and Astronomy, The University, Oxford Road,
Manchester, M13 9PL, United Kingdom}

\date{July 27, 2000}

\maketitle

\begin{abstract}
We analytically study coarsening dynamics in a system with nonconserved
scalar order parameter, when a uniform time-independent shear flow is 
present. We use an anisotropic version of the Ohta-Jasnow-Kawasaki 
approximation to calculate the growth exponents in two and three 
dimensions: for $d=3$ the exponents we find are the same as expected 
on the basis of simple scaling arguments, that is $3/2$ in the flow 
direction and $1/2$ in all the other directions, while for $d=2$ we 
find an unusual behavior, in that the domains experience an unlimited 
narrowing for very large times and a nontrivial dynamical scaling appears.
In addition, we consider the case where an oscillatory shear is applied 
to a two-dimensional system, finding in this case a standard $t^{1/2}$ 
growth, modulated by periodic oscillations. We support our two-dimensional 
results by means of numerical simulations and we propose to test our 
predictions by experiments on twisted nematic liquid crystals.
\end{abstract}

\section{Introduction}

When a statistical system in its homogeneous disordered phase is suddenly 
quenched below the critical temperature, deep into a multi-phase 
coexistence region, a dynamical process known as {\it coarsening}, or
{\it phase-ordering}, results: domains of the different 
ordered phases are formed and compete with each other in the attempt to 
break the symmetry and project the system on to one single equilibrium 
state \cite{Review}. An equivalent phenomenon occurs in the case of 
binary fluids: a system at the critical concentration tries to 
phase-separate after the quench, by forming domains of the two different 
components (spinodal decomposition).
An interesting problem is the analysis of the dynamical evolution of 
these domains, and in particular the determination of their growth rate.
In this aim a property shared by many statistical systems, called 
{\it dynamical scaling}, stating that space and time scale homogeneously 
in the equal-time two-point correlation function, $C(r,t)=f[r/L(t)]$, 
proves very useful. 
It is then natural to identify the length scale $L(t)$ as the 
typical size of the domains during coarsening. This length scale 
has generally a power-law dependence on time, $L(t)\sim t^{1/z}$, 
sometimes with logarithmic corrections. The determination 
of the exponent $z$ for many different statistical systems 
has been the object of much effort in the 
past years and we can say that ordinary coarsening 
is now quite well understood \cite{Review}.

A related topic, which is attracting growing attention in recent 
years, is the problem of phase-ordering when the system is subject 
to an external shear. Apart from the great technological relevance
of such a problem, especially in the case of spinodal decomposition,
the basic theoretical understanding of the phenomena involved  
is far from being well established \cite{OnukiRev,Beysens,Rothman}.
When a shear is present, domain growth is heavily
affected by the presence of the induced flow and the dynamical
scaling behavior is drastically different from the case of ordinary
coarsening. In particular, two main points are worthy of careful
investigation: First, the growth of the domains is anisotropic
and therefore the dynamical evolution is described by more than 
one  exponent. The determination of the shear exponents is, 
of course, of the uppermost importance. Secondly, it is not clear 
whether the shear causes an interruption of coarsening, the
dynamical balance between growth and deformation giving rise to a 
stationary state (as argued in \cite{Ohta}), or, on the contrary, 
whether domain growth continues indefinitely. 
Experimental, numerical and theoretical evidence concerning
both these points is still very tentative \cite{Cates}.

In the present work we perform a theoretical investigation of the 
coarsening dynamics in a statistical system with nonconserved scalar 
order parameter (model A, in the classification of Hohenberg and Halperin 
\cite{Hohenberg}), when a shear flow uniform in space is present. 
If, on one hand, such a model is unsuitable for describing 
spinodal decomposition in binary fluids, on the other hand it allows
us to compute the growth exponents in any spatial dimension, in 
the context of a suitably modified version of the classic 
Ohta-Jasnow-Kawasaki (OJK) approximation \cite{OJK}. When considering
the relevance of nonconserved dynamics for an advance in our
understanding of domain growth in the presence of a shear, 
we must take into account the fact that the only existing analytic 
calculations of the growth exponents for spinodal decomposition 
(conserved dynamics, or model B) have been performed in the limit of 
infinite dimension $N$ of the order parameter \cite{beppe}, where no 
saturation of coarsening is found. However, in that case the very 
concept of domains is meaningless and thus a calculation which takes 
into account the more physical case of a scalar order parameter is 
desirable. Besides, an understanding of the effect of shear on nonconserved 
coarsening is by itself an interesting problem, both from the theoretical
and experimental point of view. Indeed, experiments have been performed in 
the past on twisted nematic liquid crystals \cite{Orihara}, showing that 
these systems are a perfect test for analytical results in statistical 
models with nonconserved order parameter. Many results, from growth laws
to persistence exponents, have been successfully tested on twisted nematic
liquid crystals \cite{Yurke} and we therefore propose a shear experiment 
on such systems to check the results of our calculation.

We investigate two very different cases: in the first, a shear uniform
in time is applied and the behavior of the system is analyzed asymptotically
for very large times. In the second case, we consider a shear flow which
is periodically oscillating in time and we study the properties of the model 
for times much longer than the period of the oscillation.
The primary effect of the shear flow is naturally to stretch
the domains in the direction of the flow, such that they can be roughly 
represented as highly elongated ellipsoids, with the growth 
taking place along the main axes. Two natural length scales therefore
arise, $L_\|$ and $L_\bot$, the size of the domain along the largest 
and the smallest axes, respectively. The determination of the growth
laws for these two length scales is the main objective of this work.

In the case of a time-independent shear, our results are nontrivial and, 
especially in two dimensions, quite unexpected. 
For $d=2$ our calculation gives $L_\|\sim \gamma^{1/2} t\,(\ln t)^{1/4}$ and 
$L_\bot\sim \gamma^{-1/2}\,(\ln t)^{-1/4}$, where $\gamma$ is the shear rate,
while for $d=3$ we find $L_\|\sim \gamma t^{3/2}$ and $L_\bot\sim t^{1/2}$.
The three-dimensional exponents are the same as one would expect on the 
basis of simple scaling arguments and are compatible with calculations for 
conserved dynamics in the large-$N$ limit \cite{beppe}: the growth 
along the flow is enhanced by a factor $\gamma t$, while the 
transverse growth is unaffected by the shear. 
On the other hand, the two-dimensional result comes as 
quite a surprise: the short size of the domains $L_\bot$ goes 
asymptotically to zero for very large times, while the scale area
grows as in the unsheared case, $L_\| L_\bot\sim t$. As we shall show,
there are topological arguments supporting this last result.
As long as our approach is valid, we do not find any evidence of 
the onset of a stationary state giving rise to an interruption of the 
coarsening process. However, in two dimensions our calculation breaks 
down when the thickness of the domains becomes comparable with the 
interfacial width. We cannot say what happens when this stage is reached, 
but it is possible that some kind of stationary state occurs in this 
regime. In the case of an oscillatory shear flow in two dimensions, 
we find $L_\|\sim t^{1/2}\, \sqrt{\gamma/\omega}\,f_\parallel(t)$  
and $L_\bot \sim t^{1/2}\, \sqrt{\omega/\gamma}\,f_\perp(t)$, where
$\omega$ is the frequency of the periodic flow: both length scales 
grow like $t^{1/2}$, but are modulated by oscillatory functions, 
$f_\parallel(t)$ and $f_\perp(t)$, with the same period as the flow and 
with mutually opposite phase. In this case also, therefore, we do not find 
any stationary state.

The structure of the paper is the following. In Section II we 
introduce the OJK approximation, with the appropriate modifications
due to the presence of the shear. We end Section II by formulating some 
self-consistency equations for the matrix encoding the anisotropy of 
the domains (the {\it elongation} matrix). 
Given the
technical difficulty of such equations, in Section III we present  
some simple geometric arguments useful to achieve a better understanding
of the asymptotic behavior of the many quantities involved in the calculation.
The explicit solution of the equations in two dimensions, together with the 
calculation of the growth exponents, is carried out in Section IV 
for a time independent shear rate and in Section V for an oscillatory
shear rate, while in Section VI we solve the time-independent problem in 
three dimensions. In Section VII we present some numerical simulations in 
two dimensions, supporting our results, and in Section VIII we discuss
a possible experimental test in the context of twisted 
nematic liquid crystals. Finally, we draw our conclusions in Section IX.
A shorter account of part of this work can be found in \cite{miniorange}.

\section{The OJK approach}

The time evolution of a statistical system with nonconserved
scalar order parameter $\phi(\vec x,t)$ is described by the time-dependent
Ginzburg-Landau equation \cite{Landau}
\beq
\frac{\partial \phi(\vec x,t)}{\partial t} =
\nabla^2\phi(\vec x,t) - V'(\phi) \ ,
\eeq
where $V(\phi)$ is a double-well potential. Under the hypothesis that
the thickness $\xi$ of the interface separating different domains is 
much smaller than the size $L$ of the domains, it is possible to write
an equation for the motion of the interface itself, assumed to be 
well localized in space. 
This is the Allen-Cahn equation \cite{AC}, asserting that the 
velocity $v$ of the interface is proportional to the local 
curvature 
\beq
v(\vec x,t)=- {\vec\nabla\cdot \vec n(\vec x,t)} \ , 
	\label{allen}
\eeq
where $\vec n(\vec x,t)$ is the unit vector normal to the interface
and $\vec\nabla\cdot \vec n(\vec x,t)$ is the curvature.
The normal vector can be written in general as
\beq
{\vec n(\vec x,t)} = 
\frac{\vec\nabla m(\vec x,t)}{|\vec\nabla m(\vec x,t)|}  \ ,
	\label{enne}
\eeq
where $m(\vec x,t)$ can be any field which is zero at the interface 
of the domain, defined by the vanishing of the order parameter 
$\phi(\vec x,t)$. Given that this is the only restriction on the 
field $m(\vec x,t)$, it is convenient {\it not} to use the order 
parameter itself in order to describe the motion of the interface via
equation (\ref{allen}), but a smoother field \cite{OJK}.
Indeed, as we shall see, the principal effect of the OJK approximation
is to produce a Gaussian distribution for the field $m(\vec x,t)$, which 
would be particularly unsuitable for the highly non-Gaussian, double-peaked
distribution of the order parameter $\phi(\vec x,t)$.

From equations (\ref{allen}) and (\ref{enne}) we have
\beq
v(\vec x,t)= -\sum_{a=1}^d \frac{\partial}{\partial x_a} 
\left( \frac{\partial_a m(\vec x,t)}{|\vec\nabla m(\vec x,t)|}\right)
= - \frac{\nabla^2 m(\vec x,t)}{|\vec\nabla m(\vec x,t)|}
+ \sum_{a,b=1}^d \frac{\partial_a m(\vec x,t) \partial_b m(\vec x,t)}
{|\vec\nabla m(\vec x,t)|^2} \ 
\frac{\partial_a\partial_b m(\vec x,t)}{|\vec\nabla m(\vec x,t)|}  \ .
	\label{mod}
\eeq
By considering a frame co-moving with the interface, we can write
\beq
0=\frac{d  m(\vec x,t)}{d  t} =
\frac{\partial m(\vec x,t)}{\partial  t} 
+ \vec v_{tot}\cdot \vec\nabla m(\vec x,t) \ .
	\label{cont}
\eeq
If a shear flow is present, it can be taken into account by including
in the total velocity $\vec v_{tot}$ of the interface,  
a contribution due to velocity field $\vec u$ induced by 
the shear
\beq
{\vec v_{tot}} = v{\vec n} + {\vec u} \ , 
	\non
\eeq
where $v{\vec n}$ is the curvature driven velocity, with direction 
orthogonal to the interface and modulus given by (\ref{mod}).
By substituting relation (\ref{mod}) into equation (\ref{cont})  and
by noting that $\vec n\cdot\vec\nabla m=|\vec\nabla m|$, 
we finally get the OJK equation
\beq
\frac{\partial m(\vec x,t)}{\partial  t} + 
\sum_{a=1}^d u_a\frac{\partial m(\vec x,t)}{\partial x_a}
=\nabla^2 m(\vec x,t) - \sum_{a,b=1}^d n_a(\vec x,t)\; n_b(\vec x,t)  
\frac{\partial^2 m(\vec x,t)}{\partial x_a\partial x_b} \ .
	\label{ojk}
\eeq
This is an exact relation for the field $m(\vec x,t)$. The OJK equation
is highly nonlinear due to the dependence of the vector $\vec n$
on the field $m$ through eq.(\ref{enne}). 
The OJK {\it approximation} \cite{OJK} consists
in replacing the factor $n_a n_b$ by its spatial average
\beq
D_{ab}(t)\equiv \langle n_a(\vec x,t) n_b(\vec x,t)\rangle \ .
	\label{elo}
\eeq
Note that the {\it elongation} matrix $D_{ab}$ must satisfy the 
obvious sum rule
\beq
\sum_{a=1}^d D_{aa}(t) = 1 \ .
	\label{sum}
\eeq
In the isotropic case (${\vec u}=0$) the elongation matrix is just
$D_{ab}=\delta_{ab}/d$ by symmetry, and the OJK equation reduces to 
a simple diffusion equation with diffusion constant equal to 
$(d-1)/d$.
On the other hand, when a shear flow is present the
matrix $D_{ab}$ must encode the anisotropy induced by the shear
and it must therefore depend
on time, as the average shape of the domains does. The system of equations 
we have to solve is therefore
\beqa
\frac{\partial m(\vec x,t)}{\partial  t}  
+ \sum_{a=1}^d u_a\frac{\partial m(\vec x,t)}{\partial x_a}
&=& \nabla^2 m(\vec x,t) - \sum_{a,b=1}^d D_{ab}(t)
\frac{\partial^2 m(\vec x,t)}{\partial x_a\partial x_b}
  \ ,
	\label{pip}
\\
D_{ab}(t)&=&
\left\langle\frac{\partial_a m(\vec x,t) \; \partial_b m(\vec x,t)}
{\sum_c [\partial_c m(\vec x,t)]^2}\right\rangle  \ .
	\label{pap}
\eeqa
In this paper we will consider a space-uniform shear in the $y$ direction,
with flow in the $x$ direction. The velocity profile is therefore
given by
\beq
{\vec u}=\gamma y {\vec e}_x \ ,
	\label{flow}
\eeq
where $\gamma$ is the shear rate and ${\vec e}_x$ is the
unit vector in the flow direction. In the present Section we will 
consider the case of a time-independent shear rate $\gamma$.
A straightforward generalization of the calculation to the periodic 
case will be given in Section V.

By going into Fourier space
we can rewrite equation (\ref{pip}) as
\beq
\frac{\partial m(\vec k,t)}{\partial  t}-
\gamma k_x\frac{\partial m(\vec k,t)}{\partial k_y}=
\left(-\sum_{a=1}^d k_a^2 + 
\sum_{a,b=1}^d D_{ab}(t) k_a k_b\right) m(\vec k,t) \ .
\label{nutria}
\eeq
Note that a naive scaling analysis of the left-hand side of
this equation would give
\beq
L_x(t) \sim \gamma t\; L_y(t) \ ,
\label{naive}
\eeq
where $L_x$ and $L_y$ are the characteristic domain sizes in the $x$ 
and $y$ directions respectively. If we assume that the domain growth 
in the directions transverse to the flow is not modified by the shear, 
we obtain from (\ref{naive}) the results
\beqa
L_x(t) &\sim& \gamma t^{3/2} \label{scala} \\
L_y(t) &\sim& t^{1/2} \ , \non
\eeqa
where $L_y$ now represents any transverse direction. 
This is the simple scaling we mentioned in the Introduction.
As we shall see, result (\ref{scala}) only holds in three dimensions,
while a completely different situation occurs for $d=2$.

In order to solve equation (\ref{nutria}) 
we perform the change of variables
\beqa
q_x&=&k_x \non \\
q_y&=&k_y+\gamma k_x t \non\\
q_a&=&k_a  \ \ , \ \  \forall \ a\geq 3 
	\label{change} \\
\tau&=& t \non \ ,  
\eeqa
introducing the field $\mu(\vec q,\tau) \equiv m(\vec k,t)$.
The corresponding equation for $\mu$ reads
\beqa
\frac{\partial \ln\mu(\vec q,\tau)}{\partial \tau}&=&
-q_x^2 - (q_y-\gamma q_x\tau)^2 - \sum_{a=3}^d q_a^2 + \label{nino}
\\
&\ & D_{11}(\tau)q_x^2 +2 D_{12}(\tau) q_x (q_y-\gamma q_x\tau) + 
D_{22}(\tau)(q_y-\gamma q_x\tau)^2 + \sum_{a,b=3}^d D_{ab}(\tau) q_a q_b
\ . \non
\eeqa
The original OJK equation (\ref{ojk}), with a shear flow given 
by (\ref{flow}), is invariant under any transformation which preserves 
the sign of the product $xy$. In order to keep this symmetry, it is 
necessary for the elongation matrix $D_{ab}$ to have the following 
block-diagonal form
\beq
D_{1a}(t)=D_{2a}(t)=0  \ \ \ , \ \ \  
D_{ab}(t)=D_{33}(t)\;\delta_{ab}   \ \ \ , \ \ \
\forall  \ a,b\geq 3  \ ,
	\label{didi}
\eeq
where, to simplify the notation, we have used $D_{33}(t)$ to denote all 
the diagonal elements for $a\geq 3$.
Equation (\ref{nino}) can now be integrated to give
\beq
\mu(\vec q,\tau)= \mu(\vec q,0) \; 
\exp\left[-\frac{1}{4} \sum_{ab} q_a R_{ab}(\tau) \; q_b\right] \ ,
\eeq
with
\beqa
R_{11}(\tau)&=&4\int_0^\tau d\tau' \; 
\{[1-D_{11}(\tau')] + 2\gamma\tau' D_{12}(\tau') +
\gamma^2\tau'^2[1-D_{22}(\tau')]\} \non \\
R_{12}(\tau)&=&4\int_0^\tau d\tau' \;
\{-D_{12}(\tau') - \gamma\tau'[1-D_{22}(\tau')]\}  \non \\
R_{22}(\tau)&=&4\int_0^\tau d\tau' \;
[1-D_{22}(\tau')] \label{R} \\
R_{1a}(\tau)&=&R_{2a}(\tau)=0 \ \ ,  \ \ \forall \ a \geq 3  \non \\
R_{ab}(\tau)&=&4\delta_{ab}\int_0^\tau d\tau' \;[1-D_{33}(\tau')]
\equiv R_{33}(\tau)\; \delta_{ab}
\ \  ,  \ \ \forall \ a,b\geq 3   \non \ .
\eeqa
We can now
go back to the original field $m(\vec k,t)$, 
via the relation
\beq 
m(\vec k,t)=\mu(k_x,k_y+\gamma k_x t,k_3,\dots,k_d,t) \ ,
\eeq
to obtain
\beq
m(\vec k,t)= m(k_x,k_y+\gamma k_x t,k_3,\dots,k_d,0) \
\exp\left[-\frac{1}{4} \sum_{ab} k_a M_{ab}(t) \; k_b \right] \ , 
	\label{mm}
\eeq
with
\beqa
M_{11}(t)&=& R_{11}(t) + 2\gamma t R_{12}(t) + 
\gamma^2 t^2 R_{22}(t)  \non \\
M_{12}(t)&=& R_{12}(t) + \gamma t R_{22}(t)  \non \\
M_{22}(t)&=& R_{22}(t) \label{M} \\
M_{1a}(t)&=&M_{2a}(t)=0 \ \ ,  \ \ \forall \ a \geq 3  \non \\
M_{ab}(t)&=& R_{33}(t)\; \delta_{ab}\equiv M_{33}(t)\; \delta_{ab} 
\ \ , \ \ \forall \ a,b \geq 3  \ . \non
\eeqa
Relation (\ref{mm}) can be better understood in real space: due to the
shear flow, the field $m$ at point $(x,y,\dots)$, at time $t$, 
is the propagation 
of the initial condition at point $(x-\gamma y t,y,\dots)$.
Note that, if we assume a Gaussian distribution for  
$m(\vec k,0)$ (disordered initial condition), the field maintains
a Gaussian distribution at all the times, due to the linearity of
equation (\ref{pip}).
In order to get the correlation of $m(\vec x,t)$ 
in real space we have to average over the initial conditions
\beq
\langle m(\vec k,0)m(\vec k',0) \rangle = 
\sqrt\Delta \ \delta(\vec k+\vec k')\ .
\eeq
The equal time pair-correlation function of $m$ is therefore
\beq
C_m(\vec x,\vec x';t)\equiv \langle m(\vec x,t)m(\vec x',t) \rangle=
\sqrt{\frac{(2\pi)^{d/2}\Delta}{\det M(t)}} \
\exp\left[-\frac{1}{2} \sum_{ab} r_a [M^{-1}]_{ab}(t) \; r_b\right] \ ,
	\label{propa}
\eeq
where $r_a=x_a-x_a'$. 
All the information on the domain growth is
contained in the {\it correlation} matrix $M_{ab}(t)$. Indeed, 
the eigenvectors of $M_{ab}(t)$ give the principal elongation
axes of the domains and the square roots of its eigenvalues
give the domain sizes along these axes.

The correlation matrix is connected to the elongation 
matrix by equations
(\ref{R}) and (\ref{M}). 
In order to close the problem we have thus to 
write another set of equations, relating $M_{ab}(t)$ 
and $D_{ab}(t)$, by exploiting 
relation (\ref{pap}). If we introduce the field 
$\varphi_a(\vec x,t)\equiv \partial_a m(\vec x,t)$, we can write
\beq
D_{ab}(t)=\int {\cal D}P(\varphi) \; 
\frac{\varphi_a(\vec x,t)\varphi_b(\vec x,t)}
{\sum_c \varphi_c(\vec x,t)^2}=\frac{1}{2}
\int_0^\infty dy\; \int {\cal D}P(\varphi) \;
e^{-\frac{1}{2}y\sum_c \varphi_c(\vec x,t)^2}\; 
\varphi_a(\vec x,t)\varphi_b(\vec x,t) \ ,
\eeq
and we have thus to work out the probability distribution
${\cal D}P(\varphi)$. The field $\varphi$ is Gaussian and therefore
we just need to compute its correlator. From equation (\ref{propa}) 
we have
\beq
\langle \varphi_a(\vec x,t) \varphi_b(\vec x,t) \rangle=
\kappa \; [M^{-1}]_{ab}(t)
\ \ \ \ \ ,  \ \ \ \ \ \kappa=\sqrt{\frac{(2\pi)^{d/2}\Delta}{\det M(t)}} \ ,
\eeq
and therefore
\beq
{\cal D}P(\varphi)=\frac{1}{Z}\;
\exp\left[-\frac{1}{2\kappa}\sum_{ab} \varphi_{a}(\vec x,t) M_{ab}(t) 
\varphi_{b}(\vec x,t)\right] \, {\cal D}\varphi
\ ,
	\label{zut}
\eeq
where the constant $Z$ normalizes the distribution. By defining
\beq
N_{ab}(y,t)=M_{ab}(t)+y\;\delta_{ab} \ ,
\eeq
and by performing the rescaling $\varphi\to \varphi \sqrt\kappa$,
$y\to y/\kappa$, we can write
\beqa
D_{ab}(t)&=&\frac{1}{2}\int_0^\infty dy\;\frac{
\int {\cal D}\varphi \;
e^{-\frac{1}{2}\sum_{a'b'} \varphi_{a'}(\vec x,t) N_{a'b'}(y,t) 
\varphi_{b'}(\vec x,t)}\varphi_a(\vec x,t)\varphi_b(\vec x,t)}
{\int {\cal D}\varphi \;
e^{-\frac{1}{2}\sum_{a'b'} \varphi_{a'}(\vec x,t) M_{a'b'}(t) 
\varphi_{b'}(\vec x,t)}} \non \\
&=&\frac{1}{2}\sqrt{\det M(t)}\int_0^\infty dy\;
\frac{[N^{-1}]_{ab}(y,t)}{\sqrt{\det N(y,t)}} \ .
\eeqa
Let us introduce the following parameters in order to explicitly
write the relation above:
\beqa
\sigma(t)&\equiv& M_{11}(t)M_{22}(t)-M_{12}(t)^2=
R_{11}(t)R_{22}(t)-R_{12}(t)^2 \ , \non \\
\tau(t)&\equiv& M_{11}(t)+M_{22}(t) \label{para} \ .
\eeqa
The first equation is a particular case of the more general relation 
$\det M=\det R$, a consequence of the fact that (\ref{change}) is an 
orthogonal transformation. We can finally write
\beqa
D_{11}^{22}(t)&=&
\frac{1}{2}\sqrt{\sigma(t)M_{33}(t)^{d-2}}
\int_0^\infty dy\;\frac{M_{22}^{11}(t)+y}
{[y^2+\tau(t)y+\sigma(t)]^\frac{3}{2}[M_{33}(t)+y]^\frac{d-2}{2}} \non \\
-D_{12}(t)&=&
\frac{1}{2}\sqrt{\sigma(t)M_{33}(t)^{d-2}}
\int_0^\infty dy\;\frac{M_{12}(t)}
{[y^2+\tau(t)y+\sigma(t)]^\frac{3}{2}[M_{33}(t)+y]^\frac{d-2}{2}} 
\label{self} \\
D_{33}(t)&=&
\frac{1}{2}\sqrt{\sigma(t)M_{33}(t)^{d-2}}
\int_0^\infty dy\;\frac{1}
{[y^2+\tau(t)y+\sigma(t)]^\frac{1}{2}[M_{33}(t)+y]^\frac{d}{2}}  \ . \non 
\eeqa
Relations (\ref{R}), (\ref{M}) and (\ref{self}) 
form a closed set of equations for the 
correlation matrix $M_{ab}(t)$, or, equivalently, for the 
elongation matrix $D_{ab}(t)$. Before attempting to solve them, it is 
helpful to use physical considerations as a guide to the expected 
asymptotic form of the elongation matrix in the limit of very large 
times. To this aim, we will consider the case of a time-independent shear 
rate.

\section{Physical considerations on the elongation matrix}

When a time-independent shear flow in the $x$ direction is present, 
the domains
will be highly elongated along this direction and therefore most of 
the surface of the domains will tend to become parallel to the 
$x$ direction for very large times. We thus expect the 
following relation to hold
\beq
D_{11}(t) = \langle n_x n_x\rangle \to 0 \ \ \ , \ \ \ t\to\infty \ .
\eeq
In the two-dimensional case, due to the sum rule (\ref{sum}), 
this relation implies
\beq
D_{22}(t)= \langle n_y n_y\rangle \to 1 \ \ \ , \ \ \ t\to\infty
\ \ \ , \ \ \ d=2 \ ,
\eeq
while in dimensions $d\geq 3$ it is not {\it a priori} clear 
whether both $D_{22}$ and $D_{33}$ remain nonzero or not. The only thing
we can write is 
\beq
D_{22}(t)+(d-2)D_{33}(t)\to 1 \ \ \ , \ \ \ t\to\infty
\ \ \ , \ \ \ d\geq 3 \ .
\eeq
With regard to the off-diagonal elements of the elongation matrix,
it is not hard to convince oneself that the only nonzero ones can
be $D_{12}(t)=D_{21}(t)=\langle n_x n_y\rangle$ (see equation (\ref{didi})): 
indeed, due to the shear, the 
domains are elongated along two main axes which are {\it not} the
$(xy)$ axes, unless $t=\infty$. Therefore, the sub-matrix 
$D_{ab}^{(xy)}(t)$ cannot be diagonal for any finite time.
On the other hand, for $t\to\infty$ the two elongation axes
become coincident with $(xy)$ and thus we expect that
\beq
D_{12}(t)\to 0  \ \ \ , \ \ \ t\to\infty \ .
\eeq
It is finally clear that no qualitative difference can exist between
$d=3$ and $d>3$. Indeed, in this paper we will explicitly state the 
results only for $d=2$ and $d=3$. 

A useful exercise is to approximate a domain with an ellipsoid and 
compute the asymptotic value of $D_{ab}(t)$ as a function of the main
axes. We will do this explicitly in two dimensions and we will just
quote the main results for $d=3$.
Let us call $L_\|$ and $L_\bot$ the largest and smallest axis of a 
two-dimensional ellipse. 
Besides, let $\theta$ be the tilt angle, that is
the angle between the $x$ axis and the $L_\|$ axis (see Fig.1).
\begin{figure}
\begin{center}
\leavevmode
\epsfxsize=4in
\epsffile{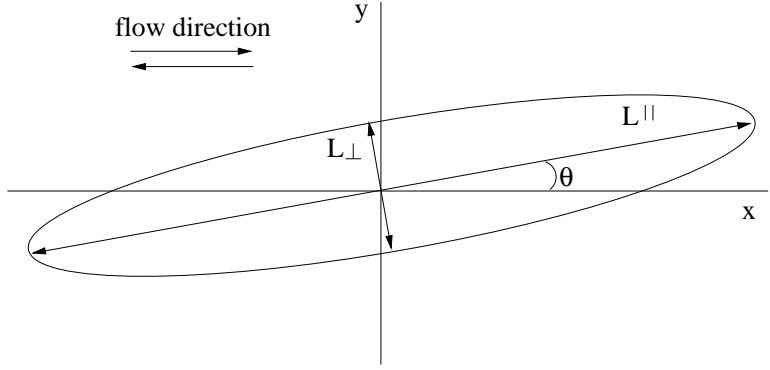}
\caption{A two-dimensional domain in the elliptic approximation.} 
\end{center}
\end{figure}
\noindent
When a time-independent shear is applied, it is natural to assume 
for $t\to\infty$:
\beq
\theta \to 0   \ \ \ , \ \ \  L_\| \gg L_\bot , 
	\label{asi}
\eeq
as an expression of the extreme elongation 
of the domain in the direction of the flow.
We can now parametrise the tilted ellipse in the following way
\beqa
x(\omega) 
&=& \frac{1}{2} L_\| \cos \omega - \frac{1}{2}\theta L_\bot \sin \omega \non
\\
y(\omega) 
&=& \frac{1}{2}\theta L_\| \cos \omega + \frac{1}{2}L_\bot \sin \omega  \ ,
\eeqa
with $\omega\in [0,2\pi]$ and 
where we have used the fact that $\theta$ is very small. The average
of any quantity $A$ along the perimeter of the ellipse can now be calculated
as,
\beq
\langle A \rangle = \frac{ \int_0^{2\pi} d\omega \, \mu(\omega) A(\omega)}
{\int_0^{2\pi} d\omega \, \mu(\omega)} \ ,
	\label{medu}
\eeq
with the metric $\mu$ given by
\beq
\mu(\omega) =\frac{1}{2} 
(L_\|^2 \sin^2 \omega + L_\bot^2 \cos^2 \omega)^{1/2} \ .
\eeq
It is useful to compute explicitly the normalizing factor in 
(\ref{medu}), i.e. the asymptotic perimeter of the ellipse,
\beq
\int_0^{2\pi} d\omega \, \mu(\omega)
=\frac{1}{2} L_\| \; \int_0^{2\pi} d\omega \,
(\sin^2 \omega + \frac{L_\bot^2}{L_\|^2} \cos^2 \omega)^{1/2} \sim
2 L_\| +  \frac{L_\bot^2}{L_\|} \ln\left(\frac{L_\|}{L_\bot}\right) \ , 
\eeq
where we have used the relation $L_\| \gg L_\bot$. The asymptotic 
perimeter divided by the total area, $L_\| L_\bot$, 
is the interfacial density $\rho$ of the domains, which must be proportional
to the energy density $E$ of the system. In the elliptic approximation
we therefore have
\beq
E \sim \rho \sim \frac{2}{L_\bot} + 
\frac{L_\bot}{L_\|^2} \ln\left(\frac{L_\|}{L_\bot}\right) \ .
\label{dusa}
\eeq
It will be interesting to compare this simple result with the 
one obtained from the OJK calculation in the next Section.

The vector normal to the interface 
can be easily found by imposing its orthogonality with the
tangent vector $(\partial_\omega x, \partial_\omega y)$. This gives
\beqa
n_x(\omega) 
&=& \frac{- \theta L_\| \sin \omega + L_\bot \cos \omega}{\mu(\omega)} 
\non \\
n_y(\omega) 
&=& \frac{ L_\| \sin \omega + \theta L_\bot \cos \omega}{\mu(\omega)} \ .
\eeqa
We can now use the relations above 
to compute the elongation matrix of the ellipse,
$D_{ab}=\langle n_a n_b\rangle$. By doing this we get
\beqa
D_{11}(t) &\sim& \theta^2 + \frac{L_\bot^2}{L_\|^2}
\ln\left(\frac{L_\|}{L_\bot}\right)
\to 0 
\non \\
D_{12}(t) &\sim& \; -\theta 
\to 0 \label{elli} \ .
\eeqa
Note that {\it a priori} we cannot say which one of the two pieces
of $D_{11}$ is going to dominate in the limit $t\to\infty$.

In dimension $d=3$ it is possible to perform a similar analysis, by 
introducing a third axis $L_z$ orthogonal to the $(xy)$ plane.
The result is
\beqa
D_{11}(t) &\sim& \theta^2 + \frac{L_\bot^2}{L_\|^2}
\to 0 
\non \\
D_{12}(t) &\sim& \; -\theta 
\to 0 \ .
\label{elli3}
\eeqa
Besides, it is possible to show that if the ratio $L_\bot/L_z$ remains
constant for $t\to\infty$, then both $D_{22}$ and $D_{33}$ are nonzero
in this limit, and
\beqa
D_{22}(t) &=& g_2(L_\bot/L_z) \non \\
D_{33}(t) &=& g_3(L_\bot/L_z) 
\label{nepo}  \ ,
\eeqa
where the two scaling functions must satisfy the relation
\beq
g_2(x)+g_3(x)=1 \ .
\eeq

The results of this Section confirm our expectation on the behavior of the
elongation matrix and also give us some hint on the relation
between the elongation matrix and the domain sizes, 
whose determination is, of course, our final goal.

\section{Time independent shear in two dimensions}

Finding a solution of the set of equations (\ref{R}), (\ref{M})
and (\ref{self}) is, even in two dimensions and with a time-independent
shear rate, not entirely straightforward. 
Therefore, we will first try to exploit a naive scaling analysis to find
a suitable ansatz for the elongation matrix,
and eventually we will
modify our initial guess in such a way to self-consistently satisfy all
our equations.

First, note that in two dimensions it is relatively simple to compute the
integrals in (\ref{self}). We obtain
\beqa
D_{11}^{22}(t)&=&\frac{\tau(t) M_{22}^{11}(t) - 2\sigma(t) - 
\sqrt{\sigma(t)}\;[2M_{22}^{11}(t)-\tau(t)]}{\tau(t)^2-4\sigma(t)}
\label{D} \\
D_{12}(t)&=&-M_{12}(t)\;\frac{\tau(t)-2\sqrt{\sigma(t)}}
{\tau(t)^2-4\sigma(t)}
\non \ ,
\eeqa
where it is easy to check that sum rule (\ref{sum}) is 
satisfied. Note that, of course, eqs.(\ref{D})  are valid 
also for a time-dependent rate $\gamma(t)$ and we will therefore 
use them also in the next Section in the case of an oscillatory shear.

A crucial task is now to understand which terms
dominate  in the limit $t\to\infty$ in the equations above.
A useful starting point is the correlation function in 
equation (\ref{propa}): if we assume that there are just 
two length scales, $L_x(t)$ and $L_y(t)$, a naive consequence we can
draw is the following
\beqa
M_{11}(t) &\sim& L_x(t)^2 \non \\
M_{12}(t) &\sim& L_x(t) L_y(t) \label{nanu} \\
M_{22}(t) &\sim& L_y(t)^2 \non \ .
\eeqa
Moreover, the physics of the system suggests that 
\beq
L_x(t) \gg L_y(t)\ .
	\label{torrone}
\eeq
Note that $L_x$ and $L_y$ {\it do not} in general coincide with 
$L_\|$ and $L_\bot$, as defined in the last Section. Indeed, this is the
main difference between the naive approach and the final full solution in 
two dimensions. Relation (\ref{torrone}) implies
\beq
M_{11}(t) \gg M_{12}(t) \gg M_{22}(t) \label{grandi}
\ , 
\eeq
and thus
\beq
\tau(t)\sim M_{11}(t) 
\ \ \ \ , \ \ \ \ 
\sigma(t)\ll M_{11}(t)^2
\label{extra}
\ .
\eeq
In order to find the asymptotic behavior of equations (\ref{D}) we need
an extra relation. From definition (\ref{para}), it seems natural to assume
that $\sigma(t)\sim M_{11}(t) M_{22}(t)$, and therefore, from
(\ref{grandi}), that
\beq
\sigma(t) \gg M_{22}(t)^2
\label{fallo}  \ .
\eeq
What we are (naively) assuming is that there are no cancellations in 
$\sigma(t)$. This assumption will fail in the final solution, but
it will only {\it logarithmically} fail, such that relation (\ref{fallo})
will still be true.
By using relations (\ref{extra}) and (\ref{fallo}) in (\ref{D}), we 
finally obtain 
\beqa
D_{11}(t)&=&\frac{\sqrt{\sigma(t)}}{M_{11}(t)}   \label{d11} \\
D_{12}(t)&=&-\frac{M_{12}(t)}{M_{11}(t)} \ , \label{d12}
\eeqa
at leading order for $t\to\infty$. Substituting relations (\ref{nanu})
into (\ref{d11}) and (\ref{d12}), and by using the naive scaling
relation $L_x(t)\sim \gamma t L_y(t)$, obtained in Section II,  
we get
\beqa
D_{11}(t)&=&\frac{L_y(t)}{L_x(t)} \sim \frac{1}{\gamma t}  \label{to1}\\
D_{12}(t)&=&-\frac{L_y(t)}{L_x(t)} \sim -\frac{1}{\gamma t} \label{to2}
\ , 
\eeqa
where again we have assumed that $\sigma(t)\sim M_{11}(t) M_{22}(t)$. 
If we now use this asymptotic
form of the elongation matrix in relations (\ref{R}) and (\ref{M}),
we obtain 
\beqa
M_{11}(t)&=& \gamma^2 t^2 \; R_{22}(t) \non \\
M_{12}(t)&=& \gamma t \; R_{22}(t) \label{nunzio} \\
M_{22}(t)&=& R_{22}(t) \non  \ ,
\eeqa
and
\beq
\sigma(t)= R_{11}(t) R_{22}(t) \ , \label{plode}
\eeq
with
\beqa
R_{11}(t)&=&4\gamma^2 \int_0^t dt' \; 
t'^2 D_{11}(t') \non \\
R_{22}(t)&=&4 \int_0^t dt' \; 
D_{11}(t') \ , \label{polo}  
\eeqa
always at leading order. 
Relations (\ref{nunzio}) 
are  consistent with equation (\ref{grandi}), and by 
substituting (\ref{nunzio}) into (\ref{d12}) we find self-consistently the 
asymptotic form $D_{12}(t)\sim -1/\gamma t$. Moreover, by assuming once
again that $\sigma(t)\sim M_{11}(t) M_{22}(t)$ and by 
substituting (\ref{nunzio}) into (\ref{d11}) we get 
$D_{11}(t)\sim 1/\gamma t$ and all our assumptions seem thus to 
be self-consistent. Unfortunately, this is not the case and it is not hard to 
understand that something is going wrong. Indeed, if we now plug into
equation (\ref{d11}) the form of $\sigma(t)$ coming from equation 
(\ref{plode}), rather than the naive assumption $\sigma(t)\sim M_{11}(t) 
M_{22}(t)$,
we get the following self-consistent equation for $D_{11}(t)$:
\beq
D_{11}(t)=
\frac{\sqrt{R_{11}(t) R_{22}(t)}}
{\gamma^2 t^2 \ R_{22}(t)}=
\frac{1}{\gamma t^2}
\left(
\frac{\int_0^t dt' \; t'^2 D_{11}(t')}{\int_0^t dt' \; D_{11}(t')}
\right)^{1/2} \ .
\label{alan}
\eeq
If we insert into the r.h.s. of this equation the asymptotic form 
of $D_{11}$ found above, we find an unpleasant surprise, that is 
\beq
D_{11}(t) = \frac{a}{\gamma t \, \sqrt{\ln \gamma t}} \ ,
\eeq
with $a=1/\sqrt{2}$, in contradiction with (\ref{to1}). 
However, the situation is far from being desperate, because if we 
try this very form of $D_{11}$ in equation (\ref{alan}) we fortunately 
find self-consistency with $a=1/2$. Our initial 
result (\ref{to1}) only failed to capture a logarithmic correction and
it is possible to check that, with this new form of $D_{11}$, we recover all
the relevant relations of this Section, namely (\ref{polo}), 
(\ref{plode}), (\ref{nunzio}), (\ref{d12}), (\ref{d11}), (\ref{fallo}),
(\ref{extra}) and (\ref{grandi}), but {\it not} (\ref{to1}).

Summarizing, the correct final form of the elongation matrix in the 
two-dimensional case is therefore (always at leading order for $t\to\infty$):
\beqa
D_{11}(t)&=& \frac{1}{2\gamma t\, \sqrt{\ln \gamma t}} \non \\
D_{12}(t)&=& - \frac{1}{\gamma t} 
\label{final} \\
D_{22}(t)&=&1-D_{11}(t) \ , \non
\eeqa
while equation (\ref{to1}) is {\it not} correct.
From (\ref{polo}) we have
\beqa
R_{11}(t) &=& \frac{\gamma \; t^2}{\sqrt{\ln \gamma t}} 
\label{rata} \\ 
R_{22}(t) &=& \frac{4\sqrt{\ln \gamma t}}{\gamma} \ , \non
\eeqa
whereas, from (\ref{nunzio}),  the correlation matrix is
\beqa
M_{11}(t) &=& 4 \gamma t^2 \, \sqrt{\ln \gamma t}  \non \\
M_{12}(t) &=& 4 t \, \sqrt{\ln \gamma t}  \label{potes} \\
M_{22}(t) &=& \frac{4\sqrt{\ln \gamma t}}{\gamma} \non \\
\sigma(t) &=& 4 t^2 \ . \non 
\eeqa
It is possible to see now that the critical assumption which went wrong
in our initial analysis was $\sigma(t)\sim M_{11}(t) M_{22}(t)$. Indeed, 
from equations (\ref{potes}) we see that $\sigma(t)$ is smaller than this, 
because there are some non-trivial cancellations in the 
determinant of $M_{ab}$. For this same reason,
one should not be misled by the fact that apparently in (\ref{potes})
the determinant
of $M_{ab}$ is null: we did not write the sub-leading contributions to
the correlation matrix which make $\sigma(t)\sim t^2 \ll t^2 \ln t$.

In order to obtain the domain size along the principal elongation
axes, $L_\|(t)$ and $L_\bot(t)$, we have to find the eigenvalues
$\lambda_1(t)$ and $\lambda_2(t)$ of $M_{ab}(t)$. 
This is easily done by recalling that the characteristic polynomial
is just $\lambda^2 -\tau\lambda + \sigma$, where $\tau$ and $\sigma$ 
are the trace and the determinant of $M_{ab}(t)$, respectively 
(cfr. eq.(\ref{para})). The final result for the two-dimensional 
case is:
\beqa
L_\|(t) &=& \sqrt{\lambda_1(t)}=\sqrt{\tau(t)} 
= 2\sqrt\gamma\;  t \ (\ln \gamma t)^{1/4} \non \\
L_\bot(t) &=& \sqrt{\lambda_2(t)} = \sqrt{\frac{\sigma(t)}{\tau(t)}}
= \frac{1}{\sqrt\gamma\; (\ln \gamma t)^{1/4}} \ . \label{end2d} 
\eeqa
Note how striking
the effect of the shear is in two dimensions: 
the size of the domains along the minor axis shrinks to zero, 
even though very slowly, for $t\to\infty$. The asymptotic effect 
of this unlimited narrowing of the domains for very large times is 
still unclear to us. However,
we do expect our approach to break down when $L_\bot(t)$ 
becomes of the same order as the interface thickness $\xi$, when 
equation (\ref{allen}) ceases to be valid. This happens after a
very large time, of the order $\exp(1/\gamma^2\xi^4)$. 
What we can say is that, if a steady state exists, it can be 
reached only when the thickness of the domains becomes comparable
with the interface width.

An important feature of the solution we have found is the failure
of standard $(x,y)$ scaling. In order to appreciate this fact, we have to
remember that, even though $L_\parallel$ and $L_\perp$ are the natural
domain sizes along the eigen-axes of the correlation matrix, other
length scales can be defined, as shown in Fig.2: 
\begin{figure}
\begin{center}
\leavevmode
\epsfxsize=4in
\epsffile{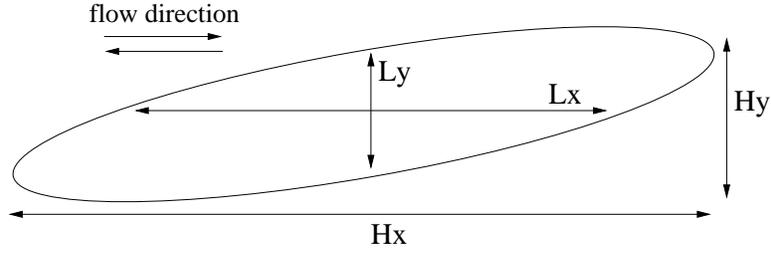}
\vskip 0.5 truecm
\caption{The length scales $L_x$, $L_y$, $H_x$ and $H_y$.} 
\end{center}
\end{figure}
\noindent
First of all, we have $L_x$
and $L_y$: from the correlation function (\ref{propa}), it follows that 
\beqa
L_x(t) &=& \frac{1}{ \sqrt{[M^{-1}]_{11}}} \non \\
L_y(t) &=& \frac{1}{ \sqrt{[M^{-1}]_{22}}} \label{lxly}
\ ,
\eeqa
and from (\ref{potes}) we get
\beqa
L_x(t) &=& \frac{\sqrt\gamma\, t}{(\ln \gamma t)^{1/4}} \non \\
L_y(t) &=& \frac{1}{\sqrt\gamma\, (\ln \gamma t)^{1/4}} 
	\label{contras}
\ .
\eeqa
Secondly, we can define $H_x$ and $H_y$, as the maximum extension of the
domain in the $x$ and $y$ directions, that is
\beqa
H_x &=& L_\parallel \cos\theta \non \\
H_y &=& L_\parallel \sin\theta  \label{hxhy} \ ,
\eeqa
where $\theta$ is the usual tilt angle (see Fig.1), which can be easily 
computed from the eigenvectors of $M$. These are
\beq
\vec e_\|=\left( 1, \frac{1}{\gamma t} \right)  \ \ \ , \ \ \ 
\vec e_\bot=\left(-\frac{1}{\gamma t}, 1 \right)  \ ,
\label{eigen}
\eeq
and therefore 
\beq
\theta =\frac{1}{\gamma t} \ .
\eeq
In this way we have
\beqa
H_x(t) &=&2\sqrt\gamma t \, (\ln \gamma t)^{1/4} \non \\
H_y(t) &=& \frac{2(\ln\gamma t)^{1/4}}{\sqrt\gamma} \ .
\eeqa 

In the absence of shear all these length scales would coincide, that is
we would have $L_\parallel=L_x=H_x$ and $L_\perp=L_y=H_y$. With the shear
this is no longer true, simply because $M_{12}\neq 0$.
Still, we would expect these length scales 
to differ only by some constant factors, 
such that they would all be of the same order asymptotically in time. 
If this situation held, we would have a standard $(x,y)$ scaling,
even though with anisotropic domains. However, in two dimensions the 
situation is very different, because the length scales above 
differ by logarithmic corrections. More precisely, we have
\beqa
L_\parallel(t) &\sim& \sqrt{\ln\gamma t} \ L_x(t) 
\sim H_x(t) \non \\
L_\perp(t) &\sim& L_y(t) \sim \frac{1}{\sqrt{\ln\gamma t}} \ H_y(t) \ .
\eeqa
The fact that $L_\parallel\neq L_x$, and therefore the emergence
of a nonstandard dynamical scaling, is closely related to the 
vanishing of the determinant of $M$ at the leading order, and 
its consequence is that $(x,y)$ are not the correct scaling axes.
We shall see that this fact does not happen in three dimensions.
In order to obtain the right scaling, we have to refer to the eigenvectors 
of the correlation matrix $M$, from which we can finally write the 
scaling form of the two-point correlation function in two dimensions  
\beq
C(x,y;t)= f\left(\frac{s}{L_\|(t)},\frac{u}{L_\bot(t)}\right) \ ,
	\label{sf}
\eeq
with
\beqa
s &=& x+y/\gamma t  \non  \\
u &=& y-x/\gamma t \label{assi}\ .
\eeqa
In the expression above, $f$ is a scaling function, while $s$ and $u$ are 
coordinates along the main scaling axes of the domains. Note that 
by $x$ and $y$ we actually mean $r_x$ and $r_y$.

Furthermore, note that the elongation matrix can be written as
\beqa
D_{11}(t) &=& \frac{L_\bot}{L_\|} \non \\
D_{12}(t) &=& -\theta \ ,
\eeqa
to be compared with the result for $D_{ab}$ obtained with the 
elliptic approximation (eq.(\ref{elli})).

An interesting quantity which can be easily computed is the 
interfacial density $\rho(t)$, defined as
\beq
\rho(t)= \langle \, \delta(m(\vec x,t))\, |\vec \nabla m(\vec x,t)| \, 
\rangle =
\int {\cal D}P(m,\varphi)\, \delta(m(\vec x,t))\, |\vec \varphi(\vec x,t)| \ ,
	\label{formula}
\eeq
where, as in Section II, we have put 
$\varphi_a(\vec x,t)=\partial_a m(\vec x,t)$. The calculation is easy 
to do because the Gaussian fields $m$ and $\varphi$ are uncorrelated. 
From relations (\ref{propa}) and (\ref{zut}) we have 
\beq
\rho(t) = \frac{\sigma(t)^{1/2}}{(2\pi)^{3/2}} \int {\cal D}\varphi \, 
e^{-\frac{1}{2} \varphi_a M_{ab} \varphi_b} \, |\vec\varphi(\vec x,t)| \ .
	\label{rho}
\eeq
By using the following formula
\beq
|\vec \varphi| = 
\frac{\int_0^\infty \frac{dy}{y^{3/2}} \left( e^{-\varphi^2 y}-1\right)}
{\int_0^\infty \frac{dy}{y^{3/2}} \left( e^{-y}-1\right) }=
\frac{1}{\Gamma(-\frac{1}{2})} 
\int_0^\infty \frac{dy}{y^{3/2}} \left( e^{-\varphi^2 y}-1\right) \ ,
\eeq
we can perform the Gaussian integral over $\varphi$ in (\ref{rho})
and, by proceeding as at the end of Section II, we get
\beq
\rho(t) = \frac{\sqrt{2}}{\Gamma(-\frac{1}{2})(2\pi)^{3/2}} 
\int_0^\infty \frac{dy}{y^{3/2}}
\left[\frac{1}{(1+y \tau/\sigma + y^2/\sigma)^{1/2}} -1 \right] 
\sim  \sqrt{\frac{\tau}{\sigma}} + 
\frac{1}{\tau} \sqrt{\frac{\sigma}{\tau}}
\ln \left(\frac{\tau^2}{\sigma}\right)
\sim \frac{1}{L_\bot} + 
 \frac{L_\bot}{L_\|^2} \ln\left(\frac{L_\|}{L_\bot}\right) \ ,
\label{densa}
\eeq
where we have used the asymptotic expressions (\ref{potes})
for $\tau(t)$ and $\sigma(t)$, together with relations (\ref{end2d}).
Remarkably, this formula for the interfacial density has the same 
asymptotic form as the one we have obtained in the context of the 
elliptic description of domains (see eq.(\ref{dusa})). 
Besides, we note an important point: $\rho(t)$ is proportional 
to the energy density of the system and therefore, given that $L_\bot(t)$ 
decreases with time (equation (\ref{end2d})), equation (\ref{densa}) means 
that the energy in the two-dimensional case {\it increases} with time
\beq
E(t) \sim  \frac{1}{L_\perp(t)} \sim \sqrt\gamma\, (\ln \gamma t)^{1/4} \ , 
\label{minkia}
\eeq
where we have subtracted the trivial ground-state contribution.
This may seem a surprising result, but we have to remember that 
due to the shear the system is not isolated, and therefore the 
dynamics is not a simple gradient descent (in other words, 
no Lyapunov functional exists). A simple example can make this 
point clearer. Imagine we prepare a two-dimensional system 
between two boundaries in a striped configuration (see Fig.2), 
with the stripes {\it orthogonal} to the boundaries 
(assume fixed boundary conditions according to the stripes). 
This configuration is stable at $T=0$. If we now shear this system,
by moving the boundaries in opposite directions, the stripes will be 
stretched and the interfacial length per unit area will increase 
(see Fig.3). Thus, in this simple case, the energy of the system 
increases under the application of a shear. This example shows that 
there is no general reason why the energy of a sheared system 
cannot increase with time. Of course, it would be important
to test equation (\ref{densa}), together with all our predictions, 
in a numerical simulation or even better in a real experiment 
(see Section VIII).
\begin{figure}
\begin{center}
\leavevmode
\epsfxsize=3in
\epsffile{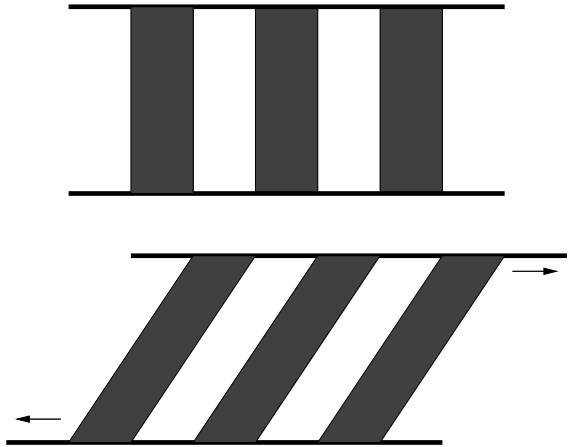}
\caption{Increasing the energy of a system by shearing it.} 
\end{center}
\end{figure}

The OJK theory also gives an explicit expression for the scaling form 
of the correlation function \cite{OJK}, which simply follows from 
equation (\ref{propa}) and from the scaling relations above
\beq
C_{OJK}(x,y;t) = \frac{2}{\pi} \sin^{-1}\left\{
\frac{\langle m(1) m(2) \rangle}
{\langle m(1)^2\rangle^{1/2}\langle m(2)^2\rangle^{1/2}} \right\}
=\frac{2}{\pi} \sin^{-1}\left\{\exp\left[-\frac{1}{2}
\left(\frac{s^2}{L_\|^2} +\frac{u^2}{L_\bot^2} \right)\right]\right\} \ .
\label{forma}
\eeq
It has been noted in \cite{beppe2} that in an unsheared, but anisotropic 
system the OJK form of the correlation function fits very well the 
numerical data. Note, however, that in the present case, 
unlike in\cite{beppe2}, the scaling laws along the two main directions 
are radically different due to the shear, and therefore it is not 
{\it a priori} clear to what extent
(\ref{forma}) is a good approximation to the scaling function $f$ in
(\ref{sf}). On the other hand, we believe that the scaling form 
we find in (\ref{sf}) has a general validity. Finally, let us note the 
elliptic symmetry of the OJK correlation function, which could explain the 
partially correct results we obtained by approximating the domains with
ellipses. The same will be true in three dimensions.

An important property of the result we have found is that
the scale area of a domain $A(t)$ satisfies the following relation:
\beq
A(t) = L_\| L_\bot = 2t \ ,
	\label{topo}
\eeq
as in the case where no shear is present. 
As we are going to explain, there are topological reasons why in two 
dimensions relation (\ref{topo}) must be satisfied either with or 
without shear. Equation (\ref{topo}) is thus a necessary 
condition fulfilled by our result, which, by itself, clearly shows
that the transverse  growth must be depressed if the longitudinal 
one is enhanced.

Let us consider an isolated domain in two dimensions in the absence 
of shear. The rate of variation of the area enclosed in the loop is 
\beq
\frac{dA(t)}{dt}=  \oint dl\; v = - \oint dl\; \vec\nabla\cdot\vec n \ ,
	\label{bebe}
\eeq
where $v$ is the velocity of the interface and $\vec\nabla\cdot\vec n$
is the local curvature (see eq.(\ref{allen})). By virtue of the 
Gauss-Bonnet theorem, the right-hand side of equation (\ref{bebe}) is 
in two dimensions a topological invariant, and therefore independent of 
the shape of the domain.

When a shear is present, we have to add to the velocity due to the 
curvature the flow velocity $\vec u$ in the direction orthogonal 
to the interface. The right-hand side of equation (\ref{bebe})
is thus corrected by the following term:
\beq
\oint dl \; \vec n\cdot \vec u = \int d^2x \; \vec\nabla\cdot\vec u =0
\ ,
	\label{bubu}
\eeq
the final equality holding for any divergence-free shear flow. 
Equation (\ref{topo}), therefore,  holds in two dimensions 
irrespective of the presence of the shear.
It is interesting that the OJK approximation, in the self-consistent
anisotropic version we have presented here, is able to capture this
essential topological feature of phase ordering in two dimensions. 
Note also that the constant $2$ in relation (\ref{topo}) is
exactly the same as one would obtain from the domain size  
in the absence of shear. We will find the same constant in the 
case of an oscillatory shear, as a further confirmation of the
validity of our method.

\section{Oscillatory shear in two dimensions}

The rather surprising results we have obtained in two dimensions
could raise the question whether the OJK method, in the modified form 
we are using here, is actually suitable for studying the physics of
a sheared system. Indeed, the skeptical reader may very well think that
the shrinking of the transverse domains size, with the consequent
increase in the total energy of the system, could be an artifact of the 
technique, rather than a genuine property of the model.
On the other hand, as we have seen at the end of the last Section, our 
two-dimensional result satisfies the highly nontrivial topological 
relation on the growth of the scale area, eq. (\ref{topo}), 
supporting the validity of our findings. Therefore, to check how robust 
our method is, we test its compatibility with the two-dimensional 
topological constraint in a completely different situation. 
To this end we study in this Section the effect of an {\it oscillatory} 
shear on phase ordering in two dimensions.

It must be said that the case of oscillatory shear is interesting 
in itself. Indeed, a realistic experimental situation is very unlikely 
to involve an indefinite time-independent shear. More reasonably,
a shear flow periodically depending on time, typically with some random 
modulation, is what we expect. Of course, real experiments with time
independent shear {\it can} be performed (and we are proposing one in 
Section VIII): what we are saying is that a generalization of
our calculation to a time-dependent oscillatory shear can shed some 
light on a more natural experimental setup.

We consider a sheared system with a velocity profile given by
\beq
\vec u = \gamma \, y \, G(t)\, \vec e_x \ ,
	\label{oscpro}
\eeq
where the only assumption we make on the shear function $G(t)$ is
that it is a periodic function with fundamental frequency $\omega$ 
and zero time average. One of the interesting aspects of the following 
calculation is that the results are to a great extent {\it independent} 
of the explicit form of $G(t)$. The derivation of 
the OJK equation is completely analogous to the one in Section II,
and it follows simply from the obvious substitution
\beq
\gamma k_x \frac{\partial m(\vec k,t)}{\partial k_y}
\ \ \ \to \ \ \ 
\gamma G(t) k_x \frac{\partial m(\vec k,t)}{\partial k_y} \ .
\eeq
In order to solve the equations we have therefore to perform the 
change of variables (compare with (\ref{change}))
\beq
q_y= k_y +\frac{\gamma}{\omega} g(t) k_x \ ,
\eeq
with
\beq
g(t) \equiv \omega \int_0^t dt' \ G(t') \ .
\eeq
Of course, for $G(t)=1$ we reproduce the time-independent shear case. 
All the equations of Section II can now be generalized to the 
oscillatory shear case by means of the following trivial 
substitution 
\beq
\gamma t \ \ \to \ \ \frac{\gamma}{\omega}\, g(t) \ . 
\eeq

A critical issue to understand concerns the regime of the parameters, 
in particular time, that we have to consider. First of all, we cannot 
afford to have too high a frequency, otherwise there would be a delay 
in the response of the system to the shear. This means we must take 
the shearing frequency, $\omega$, much smaller than the shear rate, 
$\gamma$. On the other hand, we need to observe the system on time 
scales much larger than a period. Therefore, we will consider the following 
regime
\beq
\frac{1}{\gamma} \ll \frac{1}{\omega} \ll t \ ,
	\label{regime}
\eeq
which implies
\beq
\alpha \equiv \frac{\gamma}{\omega} \gg 1 \ .
	\label{alpha}
\eeq
Note that, in this way, we cannot recover from our final results
the $\omega\to 0$ case, nor can we extrapolate to the $\omega \to\infty$ 
limit. On the other hand, the large parameter $\alpha$ will be useful for
extracting the leading terms from our results.

Before going further, let us explain our general strategy. Due to the 
periodic shear, all our quantities will exhibit oscillations: some
of them, like $D_{11}$, which is positive definite, will oscillate
around a nonzero value, while others, like $D_{12}$, will oscillate 
around zero, due to the oscillation in the orientation of the domains. 
Given that all these quantities enter the time integrals in equations
(\ref{R}), a natural approach, for times much longer than the 
period, is to exploit their time-average: if $B(t)$ is an oscillatory 
quantity we write, to leading order for $t \to \infty$, 
\beq
\int_0^t dt' \ B(t') \ t'^n \sim \bar B \ t^{n+1} \ ,
\eeq
with
\beq
\bar B = \frac{\omega}{2\pi} \int_0^\frac{2\pi}{\omega} dt' \ B(t') \ .
\eeq
In this way from equations (\ref{R}) we get
\beqa
R_{11}(t) &=& 4t \left\{ 1 - \overline{D_{11}} + 
                        2 \alpha \ \overline{g D_{12}} 
                      + \alpha^2 \ \overline{g^2 D_{11}} \right\} \non \\
R_{12}(t) &=& 4t \left\{ -\overline{D_{12}} - \alpha 
                 \ \overline{gD_{11}} \right\} \\
R_{22}(t) &=& 4t \ \overline{D_{11}} \non \ . 
\eeqa
Note the striking difference from the time-independent shear case: due 
to the oscillations the whole matrix $R_{ab}$ is now of order $t$, as 
it would be in the absence of shear. As mentioned above, we expect $D_{12}$ 
to oscillate around zero with the same period as $g$. Thus, in the equations 
above we can disregard terms like $\overline{D_{12}}$ and $\overline{gD_{11}}$,
whose time average is zero. As a consequence, we have
\beq
R_{12}(t) = 0 \ ,
\eeq
that is, the isotropy is restored at the level of the matrix $R_{ab}$.
On the other hand, we have to keep mixed terms like $\overline{g D_{12}}$,
because their time average will be nonzero. Using equations (\ref{M})
we can now write
\beqa
M_{11}(t) &=& 4t \left\{ 1 - \overline{D_{11}} + 
                        2 \alpha \ \overline{g D_{12}} 
                      + \alpha^2 \ \overline{g^2 D_{11}} 
                      + \alpha^2 g^2(t)\ \overline{D_{11}}
\right\} \non \\
M_{12}(t) &=& 4t \  \alpha \ g(t)\  \overline{D_{11}} 
\label{spanu} \\
M_{22}(t) &=& 4t \ \overline{D_{11}} \non \ . 
\eeqa
First of all note that, apart from the oscillation induced by the explicit 
presence of $g(t)$, the correlation matrix $M_{ab}$ is of order $t$, strongly
suggesting that we will end up with a $t^{1/2}$ growth. On the other hand,
$M_{12}\neq 0$, meaning that the system is still anisotropic, even though
the anisotropy has zero time average. 

From relation (\ref{alpha}) and from equations (\ref{spanu}), we have
that $M_{11} \gg M_{12} \gg M_{22}$, and therefore the self-consistent
equations (\ref{D}), become
\beqa
D_{11}(t) &=& \frac{M_{22}(t) + \sqrt{\sigma(t)}}{M_{11}(t)} \non \\
D_{12}(t) &=& - \frac{M_{12}(t)}{M_{11}(t)} \label{gino} \ .
\eeqa

As usual, we need a starting point to break into these equations and
some physical considerations may help here. First, note that naively 
$M_{12}\sim L_x L_y \sim t$, from the topological relation (\ref{topo}).
The second of equations (\ref{spanu}) then suggests that 
$\overline{D_{11}}\sim 1/\alpha$. Secondly, from the form of the 
velocity profile, we have another naive relation, that is 
$x \sim \alpha \, y\,  g(t)$. Thus, we expect that $-D_{12}\sim \theta
\sim y/x \sim 1/\alpha$. We therefore make the following ansatz
\beqa
D_{11}(t) &=& \frac{1}{\alpha} f(t) \non \\
-D_{12}(t) &=& \frac{1}{\alpha} h(t) \ ,
\eeqa
with $f(t)\ge 0$, while we expect $h$ to oscillate around zero. 
Both $f$ and $h$ must now be determined 
self-consistently. Inserting this ansatz into (\ref{spanu}) and
considering only the leading terms in $\alpha$, we have
\beqa
M_{11}(t) &=& 4 t \, \alpha [u + r\, g^2(t)] \non \\
M_{12}(t) &=& 4t \, g(t) r \non \\
M_{22}(t) &=& 4t \, \frac{r}{\alpha} \label{nonse} \\
\sigma(t) &=& 16 t^2 \, r u \ ,  
\eeqa
with the two constants $u$ and $r$ given by
\beq
r\equiv \overline{f}  \ \ \ \ \ \ \ \ \ \ \ \ \  u\equiv\overline{g^2 f} \ .
\eeq
By inserting this form of $M_{ab}$ into equations (\ref{gino}), we find
that the powers of $\alpha$ balance and we obtain two equations for 
the functions $f$ and $h$
\beqa
f(t) &=& \frac{\sqrt{ru}}{u+r g^2(t)} \label{prima} \\
h(t) &=& \frac{r g(t)}{u+r g^2(t)} \ . \non 
\eeqa
Averaging the first equation, we get
\beq
\sqrt{ru}= \Omega(r/u) \ ,
\eeq
with
\beq
\Omega(x) = \overline{\left\{\frac{1}{1+x g^2(t)}\right\}} \ .
\eeq
On the other hand, by multiplying the same equation by $g^2(t)$ and
averaging again, we have
\beq
\sqrt{ru} = 1 - \Omega(r/u) \ ,
\eeq
and therefore
\beq 
\sqrt{r u}=\Omega(r/u)=1/2 \ .
	\label{ultracheck}
\eeq
In order to compute the domain sizes we can use the same formulae as in 
Section IV, because we still have $\tau^2 \simeq M_{11}^2\gg \sigma$. We 
obtain
\beqa
L_\parallel(t) &=& \sqrt{\tau} = 2 t^{1/2} \ \sqrt{\frac{\gamma}{\omega}}
\ \sqrt{u +r g^2(t)} \\
L_\perp(t)  &=& \sqrt{\frac{\sigma}{\tau}} = 2  t^{1/2} \
\sqrt{\frac{\omega}{\gamma}}
\ \frac{\sqrt{ru}}{\sqrt{u +r g^2(t)}} \ ,
\eeqa
and, happily, we find for the scale area
\beq
A(t) = L_\parallel(t) L_\perp(t) = 4 t \sqrt{ru} = 2 t \ ,
\eeq
{\it independent} of the explicit form of the shear function $G(t)$.
Note also that the factor $2$ in this formula is exactly the same 
as in the time-independent shear case and in the unsheared case.
This is an important result, supporting the validity of our method 
for the study of the effect of shear in this type of system.

As expected, apart from the oscillations, the growth follows 
a $t^{1/2}$ law. The interesting thing is that both $L_\parallel$ and 
$L_\perp$ oscillate in time, but, as expected, with an opposite phase:
when $g(t)$ has its maximum (i.e. at the maximum shear displacement),
$L_\parallel$ is maximum and of course $L_\perp$ is minimum, because 
this is the point of maximum elongation of the domains. 
On the other hand, for $g(t)=0$ (i.e. zero shear displacement) $L_\parallel$
is minimum and $L_\perp$ maximum, but always with $L_\parallel \gg L_\perp$.
We want to stress that this oscillatory dynamics is only deceptively simple. 
To better appreciate this fact we have to compute $L_x$ and $L_y$ 
(see Fig.2 and eq.(\ref{lxly})). These quantities read
\beqa
L_x(t)&=&2 t^{1/2} \ \sqrt{\frac{\gamma}{\omega}}
\ \sqrt u \\
L_y(t)&=&2 t^{1/2} \ \sqrt{\frac{\omega}{\gamma}} 
\frac{\sqrt{ru}}{\sqrt{u+r g^2(t)}}
\ .
\eeqa
First of all, note that $L_x$, unlike $L_y$, does {\it not} oscillate in 
time, and this had to be expected from its very definition (see Fig.2).
Secondly, note that for $g(t)=0$, we have $L_x=L_\parallel \gg L_y=L_\perp$:
at the points of zero shear displacement the domains are very flat 
and large. Besides, we can compute the tilt angle $\theta$ from the 
eigenvectors of $M$, thus obtaining
\beq
\tan\theta = \frac{\omega}{\gamma} \
\frac{r g(t)}{u+r g^2(t)}  \label{tilt} \ .
\eeq
We can see that $\theta$ is zero at the zero displacement point ($g(t)=0$),
and increases with increasing displacement, up to a maximum, whose
value decreases with increasing shear rate $\gamma$. This fact may seem 
counterintuitive, especially because in the case of a time-independent 
shear rate we have seen that the tilt angle was {\it decreasing} with time, 
while here it is {\it increasing}.
However, there is no contradiction: the key point is that at $g(t)=0$ 
the domains are {\it already} very elongated, that is $L_x\gg L_y$, as 
an effect of the shear experienced in the former periods. We can better 
understand what happens by using the simple case of a linearly sheared 
ellipse (no growth), with initial axes $L_x$ and $L_y$, and  
$L_x \gg L_y$. The ellipse is described by the parametric equation 
\beqa
x&=&L_x \cos\phi +\gamma y t \non \\
y&=&L_y \sin\phi \ ,
\eeqa
with $\phi\in[0,2\pi]$. We can estimate the tilt angle by
computing the ratio $y/x$ at the point where the $x$ displacement is 
maximum. This gives
\beq
\tan\theta = \frac{ \gamma t}{\gamma^2 t^2 + L_x^2/L_y^2} \ .
\eeq
This function has a maximum at
\beq
t_{\rm max}= \frac{1}{\gamma} \ \frac{L_x}{L_y} \ ,
\eeq
and decreases asymptotically like $1/\gamma t$ for $t\to\infty$.
In the case of time-independent shear rate, the initial condition $t=0$ has
$L_x/L_y\sim 1$, and therefore the maximum of $\theta$ is quickly reached 
at $t_{\rm max} \sim 1/\gamma$, which is much smaller than the times we
consider, $t\gg 1/\gamma$. For this reason, in the time region of interest 
the tilt angle monotonically decreases. In the oscillatory shear case, 
on the other hand, at the zero displacement point, $g=0$, we have 
$L_x/L_y\sim \gamma/\omega$ and thus $t_{\rm max}\sim 1/\omega$: 
the tilt angle therefore increases during the period of the oscillations,
and this explains the apparent contradiction between the two cases.

From the tilt angle (\ref{tilt}) we can compute the additional length 
scales $H_x$ and $H_y$ by using definition (\ref{hxhy}). We have
\beqa
H_x(t)&=&2 t^{1/2} \ \sqrt{\frac{\gamma}{\omega}}
\ \sqrt{u +r g^2(t)}  \\
H_y(t) &=& 2 t^{1/2} 
\sqrt{\frac{\omega}{\gamma}} \frac{r g(t)}{\sqrt{u +r g^2(t)}}
\ .
\eeqa
Note that $H_y(t)$ is the only length scale to vanish at the zero 
displacement point. After the discussion above, the reason for this should 
now be clear.

In order to compute $r$ and $u$, we need to know the explicit form of
$g(t)$, and therefore of $G(t)$. However, these are just numerical constants
and the time evolution of the domain sizes is not affected by them.
For the particularly simple case where
\beq
G(t)= \cos \omega t \ ,
\eeq
the constants are
\beq
r=\frac{\sqrt 3}{2} \ \ \ \ \ \ \ \ u=\frac{1}{2\sqrt 3} \ .
\eeq

\section{Time independent shear in three dimensions}

In dimension larger than two it becomes very difficult to 
explicitly compute the integrals in equations (\ref{self}).
Notwithstanding this, if we formulate a suitable ansatz for
the elongation matrix $D_{ab}(t)$, we can then find $M_{ab}(t)$
from equations (\ref{R}) and (\ref{M}), and finally obtain a 
self-consistent relation for $D_{ab}(t)$ by an asymptotic evaluation
for large times of the integrals in (\ref{self}). In the present
Section we will carry out this program for a time-independent shear 
rate.

First of all, we note that many of the terms in equations (\ref{R}) can
be estimated by means of the following reasonable ansatz:
\beqa
D_{11}(t) &\to& 0 \ \ \ , \ \ \ t\to\infty \non \\
D_{12}(t)&\sim&-\frac{1}{\gamma t} \ .
\label{simple}
\eeqa
Both these relations are also obtained in any dimension by the 
calculation of Section III. By inspection of equations (\ref{R}) 
it is now clear that the key quantity needed to evaluate $R_{ab}(t)$, 
and thus $M_{ab}(t)$, is $[1-D_{22}(t)]$.
We could be tempted to try an ansatz similar to the case $d=2$, by taking
$[1-D_{22}(t)]\sim (\ln \gamma t)^{a_1}/t^{a_2}\to 0$, for $t\to\infty$. 
However, a careful analysis of the equations shows that this ansatz is 
not consistent. Therefore, the most natural thing to do is to assume 
that {\it both} $D_{22}(t)$ and $D_{33}(t)$ remain nonzero for
$t\to\infty$, that is (according to the usual sum rule)
\beqa
D_{22}(t)&\to& 1-K \label{sgau} \\
D_{33}(t)&\to& K \non \ , 
\eeqa
and to fix self-consistently the value of the constant $K$. From 
equations (\ref{R}), (\ref{M}), (\ref{para}) and (\ref{sgau}) we have
\beqa
M_{11}(t) &=& \frac{4}{3} K \gamma^2 t^3 \non  \\
M_{12}(t) &=& 2 K \gamma t^2  \non \\
M_{22}(t) &=& 4 K t  \label{tredi} \\
M_{33}(t) &=& 4 (1-K) \; t \non \\
\sigma(t) &=& \frac{4}{3} K^2 \gamma^2 t^4 \ , \non  
\eeqa
at leading order for $t\to\infty$. Note that the explicit forms of
$D_{11}(t)$ and $D_{12}(t)$ do not enter in $M_{ab}(t)$. 
Using relations (\ref{tredi}) it is now possible to evaluate the
asymptotic value of the integrals in equations (\ref{self}) and get an 
equation for the constant $K$. 
In three dimensions equations (\ref{self}) read   
\beqa
D_{22}(t) &=&
\frac{1}{2}\sqrt{\sigma(t)M_{33}(t)}
\int_0^\infty dy\;\frac{y^2+M_{11}(t)\,y+M_{11}(t)M_{33}(t)}
{\{y^3+ M_{11}(t)\, y^2 + 
[M_{11}(t)M_{33}(t)+\sigma(t)]\; y +\sigma(t)M_{33}(t)\}^{3/2}}
\non \\
D_{33}(t) &=&
\frac{1}{2}\sqrt{\sigma(t)M_{33}(t)}
\int_0^\infty dy\;\frac{y^2+M_{11}(t) \,y+\sigma(t)}
{\{y^3+ M_{11}(t)\, y^2 + 
[M_{11}(t)M_{33}(t)+\sigma(t)]\; y +\sigma(t)M_{33}(t)\}^{3/2}} \ , \non
\eeqa
where we have used the relation $M_{11}(t) \gg M_{22}(t)\sim M_{33}(t)$, 
according to (\ref{tredi}). 
By performing the rescaling $y\to t y$ and by using relations (\ref{tredi})
in the two integrals above, it is possible to see that in the limit 
$t\to\infty$ we can disregard the terms $y^2$ at the numerator 
and $y^3$ at the denominator. 
In this way we obtain
\beqa
D_{22}(t) &=&
\frac{\sqrt\alpha}{2}
\int_0^\infty dy\;\frac{y + 1 -\beta}
{(y^2+ y + \alpha)^{3/2}} \non \\
D_{33}(t) &=&
\frac{\sqrt\alpha}{2}
\int_0^\infty dy\;\frac{y+\beta}
{(y^2+ y + \alpha)^{3/2}} \label{into} \ ,
\eeqa
with
\beqa
\alpha &=& \frac{4K(1-K)}{(4-3K)^2} \non \\
\beta &=& \frac{K}{4-3K} \label{ab} \ .
\eeqa
The fact that there is no time-dependence left in the right-hand sides
of equations (\ref{into}) 
shows that ansatz (\ref{simple}) and (\ref{sgau}) give rise to
a self-consistent solution for the three-dimensional case.
Moreover, it is straightforward to check that sum rule (\ref{sum}) is
satisfied. The integrals in (\ref{into}) can now be easily performed 
and, by using relations (\ref{sgau}), after some algebra we find 
\beq
K=1/5 \ .
	\label{kappa}
\eeq
A similar treatment of the integrals in (\ref{self}) for
$D_{11}(t)$ and $D_{12}(t)$ shows that,
\beqa
D_{11}(t) &\sim& \frac{\ln(\gamma t)}{\gamma^2 t^2} \non \\
D_{12}(t) &\sim& -\frac{1}{\gamma t} \non
\eeqa
consistent with ansatz (\ref{simple}).
Let us note that relations (\ref{simple}), (\ref{sgau}) and
(\ref{tredi}) are self-consistent in any dimension $d\geq 3$, as can 
be easily verified by using these relations in equations (\ref{self}) 
and rescaling $y\to t y$ in the integrals. Our final result will 
therefore be qualitatively the same for any dimension $d \ge 3$ (for 
$d>3$ only numerical factors, such as the values of $K$ and the 
amplitudes in equations (\ref{end3d}) below, are changed). 

We can now compute the eigenvalues of the correlation matrix $M_{ab}(t)$, 
in order to find the sizes of the domains along the principal elongation 
axes. From (\ref{tredi})
and (\ref{kappa}), we have
\beqa
L_{||}(t) &=& \frac{2}{\sqrt{15}} \; \ \gamma\;  t^{3/2} \non \\
L_\bot(t) &=& \frac{1}{\sqrt{5}} \;   \      t^{1/2} \label{end3d} \\
L_z(t) &=& \frac{4}{\sqrt{5}} \;   \      t^{1/2} \non \ ,
\eeqa
whose corresponding eigenvectors are
\beq
\vec e_\|=\left( 1, \frac{3}{2\gamma t}, 0 \right) \ \ , \ \ 
\vec e_\bot=\left(-\frac{3}{2\gamma t}, 1, 0 \right) \ \ , \ \ 
\vec e_z=\left( 0, 0, 1 \right) \ ,
\eeq
where we recall that $L_{||}$ and $L_\bot$ are the larger and smaller
orthogonal axes of the domain in the $(xy)$ plane, whereas $L_z$
is the axis of the domain in $z$ direction (or any direction orthogonal 
to the $(xy)$ plane, if $d>3$). The domain growth in dimension $d \ge 3$ 
is therefore the one we would expect on the basis of the simple scaling 
arguments given in Section II (see equations (\ref{scala})): the 
growth exponent along the flow direction is augmented by one, whereas
the others are left unchanged. Unlike the two-dimensional case, there 
are no topological restrictions on the product of the domain sizes, 
because the integral over the domain surface of the local curvature
is not, in $d\neq 2$, a topological invariant.

As already anticipated, for $d=3$ standard scaling holds. Indeed, one 
can immediately check that
\beqa
L_\parallel(t) &\sim& L_x(t) \sim H_x(t) \non \\ 
L_\perp(t) &\sim& L_y(t) \sim H_y(t) \ . 
\eeqa
There is therefore no real difference between growth along the 
principal axes of the domains and growth in the $(xyz)$ directions, 
and the correlation function displays the simple asymptotic scaling form
\beq
C(x,y,z;t) =f\left( \frac{x}{L_\|}, \frac{y}{L_\bot}, \frac{z}{L_z}\right) \ .
\label{calmo}
\eeq
According to the OJK theory \cite{OJK}, we have, for $t \to \infty$, 
\beq
C_{OJK}(x,y,z;t) =\frac{2}{\pi} \sin^{-1}\left\{\exp\left[-\frac{1}{2}
\left(\frac{(x+3y/2\gamma t)^2}{L_\|^2} +\frac{(y-3x/2\gamma t)^2}{L_\bot^2} 
+ \frac{z^2}{L_z^2}
\right)\right]\right\} \ .
\label{forma3}
\eeq
In the scaling limit, where $x,y,z,t \to \infty$ with $x/L_\|$, $y/L_\bot$ 
and $z/L_z$ fixed, the term $3y/2\gamma t$ can be dropped, but the term 
$3x/2\gamma t$ cannot, and the OJK scaling function has ellipsoidal symmetry 
as expected.   

As in two dimensions, we can compute the interfacial density by applying 
equation (\ref{formula}). The final result is
\beq
\rho(t) \sim \frac{1}{L_\bot} + \frac{L_\bot}{L_\|^2} \ ,
\eeq
which shows that the energy density in the three dimensional case decreases
in the standard way, 
\beq
E(t) \sim t^{-1/2} \ . 
\eeq

\section{Numerical simulations in two dimensions}

In the present Section we will present some numerical simulations for 
a two-dimensional system subject to a time-independent uniform shear. 
We have considered a system of Ising spins on a lattice,  governed by 
zero-temperature Monte Carlo dynamics. As in the rest of this paper, 
the shear flow is applied in the $x$ direction, according to the profile 
given by equation (\ref{flow}). From a practical point of view, we have 
sheared the system by shifting each row of spins by an amount 
proportional to the $y$ coordinate and to the time $\hat t$ (measured 
in Monte Carlo steps)
\beq
\Delta x(y,\hat t) = y\,n_s(\hat t)  \ ,
\eeq
where $n_s(\hat t)$ is the number of discrete shear steps up to time 
$\hat t$. Of course, the discrete nature of the system is reflected in 
the discrete nature of the shearing process. To simulate a shear rate 
$\gamma$ (defined by $\Delta x_{\rm continuous} = \gamma y \hat t)$) we 
require $n_s(\hat t) = {\rm Int}\,(\gamma \hat t)$: After each $1/\gamma$ 
Monte-Carlo steps a discrete shear process, where each row moves one 
lattice spacing relative to the row below it, is applied. In the 
large-time limit, where $\hat t \gg 1/\gamma$, the system's behavior 
should not be very different from that of a continuously sheared system.  

We have to be careful in choosing the boundary conditions for a sheared 
system, because normal periodic boundary conditions would clearly be 
wrong. The idea is to replicate the original system infinitely many times 
on the $(x,y)$ plane and to shear each sub-system with respect to the 
others. In other words, if $(x,y)$ are the coordinates on the infinite 
plane, and $(i,j)$ are the coordinates on our numerical system, we have
\beqa
i &=& {\rm Mod}_{N_x}(x + y\,n_s(\hat t)) \non \\
j &=& {\rm Mod}_{N_y}(y) \ ,
	\label{pbc}
\eeqa
where $N_x$ and $N_y$ are the sizes of the numerical system in the $x$ and
$y$ directions, and the function ${\rm Mod}_N(z)$ is just the value of $z$ 
modulo $N$. Clearly, for $\hat t=0$ equations (\ref{pbc}) reduce to standard 
periodic boundary conditions.

One of the main difficulties in simulating a system subject to a shear is that
the domains grow very quickly in the direction of the flow, soon reaching 
a size comparable with the size of the system. On the other hand, as we have
seen, we expect the growth to be highly depressed in the transverse direction.
Thus, the most reasonable thing to do is to take $N_x \gg N_y$, in order to
reduce as much as possible finite size effects. In all our simulation we have
taken $N_x=20000$ and $N_y=100$. As we shall see, even for our longest times,
the domains are much smaller than the size of the system in both directions.
A possible proposal in order to reduce the finite size effects due to the 
shear-induced elongation of the domains is to work at very low $\gamma$.
However, all our results hold in the limit $L_x \gg L_y$: if we decrease
the shear we will have to wait for a longer time to enter the asymptotic 
regime of interest, and thus we will still have the problem of long 
domains compared to the system size. There is, therefore, no easy way out 
of this situation and we had to tune our parameters to take this problem 
into consideration.
For this reason we run our simulations for only one value of the
shear rate, namely $\gamma=1/4$: in order to study the dependence of 
all the observables on the shear rate we would have to consider values 
of $\gamma$ far from the suitable numerical domain.

The first thing we want to check is the behavior of the length scales
$L_x(t)$ and $L_y(t)$. As we have seen, $x$ and $y$ are not the correct
scaling axes, but we want to test our prediction for $L_x$ and $L_y$ 
against the naive expectation of equation (\ref{scala}). Indeed, it must
be remembered that this naive scaling is also the one found in the 
case of conserved dynamics in the limit of infinite dimension 
of the field \cite{beppe}. 
We recall our analytic prediction
\beqa
L_x(t) &\sim& \frac{\sqrt\gamma\, t}{(\ln \gamma t)^{1/4}} \non \\
L_y(t) &\sim& \frac{1}{\sqrt\gamma\, (\ln \gamma t)^{1/4}} \sim L_\perp(t)
	\label{contras2}
\ .
\eeqa
Note that $L_y$ is, at the leading order, equal to $L_\perp$, and therefore
we can limit ourselves to measure the former. This is important, because
a numerical measure of $L_\perp$ would be very difficult: the domain size 
in the perpendicular direction is very small and for long times this
direction passes through very few lattice sites, such that practically
there are no points where the correlation function is different from zero. 
This problem does not exists for the correlation in the $x$, $y$ and parallel
direction. In order to extract the the domain scale at a given time we have 
performed a fit of the correlation function to the OJK form and have 
located the point where the fit is equal to $1/e$. We have checked that the 
behavior of the domain size with time is almost entirely insensitive to the 
particular fit we use. Numerically, we do not expect to be able to detect 
the logarithmic corrections in (\ref{contras2}), so our goal is to check the 
leading behavior $L_x\sim t$ and $L_y\sim O(1)$. 
Our results are shown in Fig.4.
\begin{figure}
\begin{center}
\leavevmode
\epsfxsize=4in
\epsffile{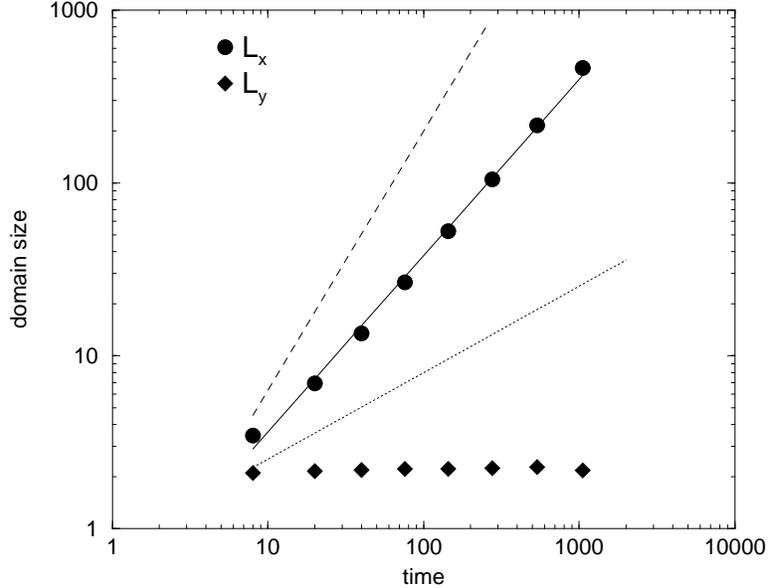}
\caption{The domain sizes in the $x$ and $y$ direction as a function 
of time. The full line is a power fit giving $L_x\sim t^{1.02}$.
The dashed line is $L_x^{\rm naive}\sim t^{3/2}$, and the dotted line
is $L_y^{\rm naive}\sim t^{1/2}$, for a comparison.
In both cases, the data are averaged over $5$ samples.} 
\end{center}
\end{figure}
As we can see, $L_x$ is definitely not growing like $t^{3/2}$.
A power fit gives
\beq
L_x(t)\sim t^{1.02} \ .
\eeq
Furthermore, $L_y$ is, on this scale, compatible with a constant, and 
is certainly not growing like $t^{1/2}$. Both $L_x(t)$ and $L_y(t)$ 
have the expected behavior, apart from the logarithmic corrections,
and the naive exponents $3/2$ and $1/2$ are clearly not correct.

The value of $L_y$ is very small and in order to have a better idea
of the fast decay of the correlation in the $y$ direction, we plot
in Fig.5 the correlation function and the OJK fit for a given fixed value
of the time. Note that actually the correlation vanishes on average
after six lattice spacings.

The next important quantity we want to measure is the energy. From 
relation (\ref{minkia}) we can see that $E(t)$ is a direct measure
of $L_\perp(t)$. Note that the relation between $E$ and $L_\perp$ is,
at leading order, completely independent on the OJK approximation
we are using: indeed, the simple assumption $L_\parallel \gg L_\perp$
is sufficient to conclude that, at leading order, $E\sim 1/L_\perp$. 
However, we stress that the condition $L_\parallel \gg L_\perp$ is 
only satisfied for large times (see Fig.4).
In Fig.6 we plot the energy as a function of time, both for the sheared 
and the unsheared case. We see that, after an initial drop in the time 
regime where we do not expect relation (\ref{minkia}) to hold, the energy 
becomes compatible with a constant on this scale. The difference with the 
unsheared case is striking. In the inset of this figure we show a 
magnification of the last part of the curve for the sheared case: 
it is encouraging to see that, despite the significantly large error bars, 
an increase in the energy for very large times is clearly visible, 
compatible with our analytic prediction
\beq
E(t)\sim \sqrt\gamma \, (\ln \gamma t)^{1/4} \ .
\eeq 
However, we stress that longer simulational times and larger system sizes 
are necessary to test this prediction (in particular, the power of the 
logarithm) more carefully. 
\begin{figure}
\begin{center}
\leavevmode
\epsfxsize=4in
\epsffile{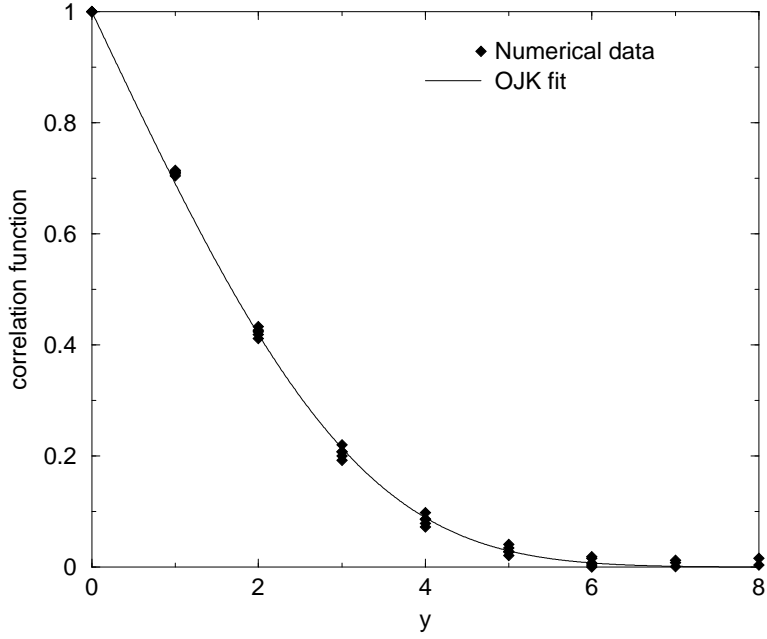}
\caption{Correlation function in the $y$ direction as a function of $y$, 
for $\hat t=144$. The symbols are the numerical data for $5$ samples; the 
full line is the OJK fit.} 
\end{center}
\end{figure}

\begin{figure}
\begin{center}
\leavevmode
\epsfxsize=4in
\epsffile{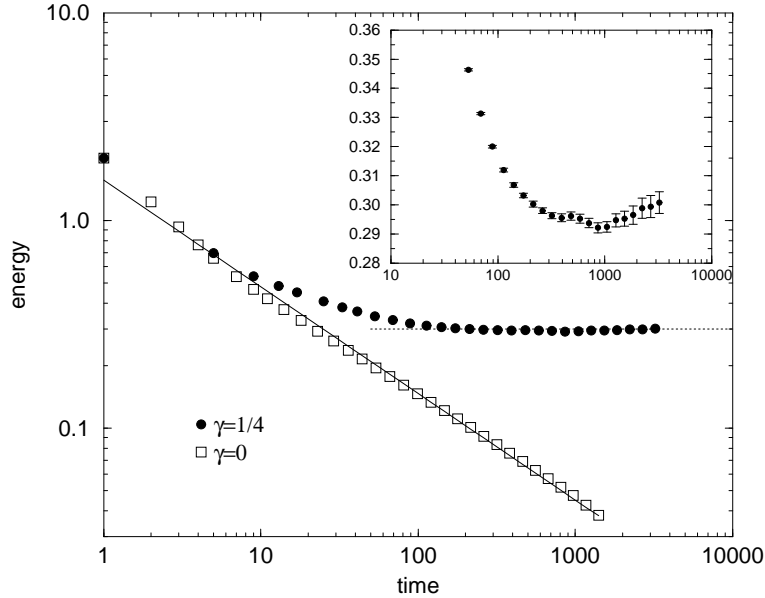}
\caption{The energy as a function of time in the sheared case ($\gamma=1/4$), 
averaged over 41 samples, and the unsheared case ($\gamma=0$) averaged over 
5 samples. The full line is a power fit for the unsheared case, giving 
$E_{\rm un}\sim t^{-0.51}$. The horizontal broken line is a guide to the eye. 
Inset: energy in the sheared case as a function of time (magnification).
} 
\end{center}
\end{figure}
The last quantity we measure is $L_\parallel(t)$, whose form 
(\ref{end2d}) differs from that of $L_x(t)$ only by a logarithmic 
correction. 
In Fig.7 we plot $L_\parallel$ as a function of the time.
Even if slightly faster, the growth of the domains in the 
parallel direction is compatible with $t$. Indeed, a power fit gives
\beq
L_\parallel(t) \sim t^{1.14} \ .
\eeq
Not surprisingly, at a simulational level we are unable to detect 
any significant difference between the growth of $L_x$ and $L_\parallel$.
\begin{figure}
\begin{center}
\leavevmode
\epsfxsize=4in
\epsffile{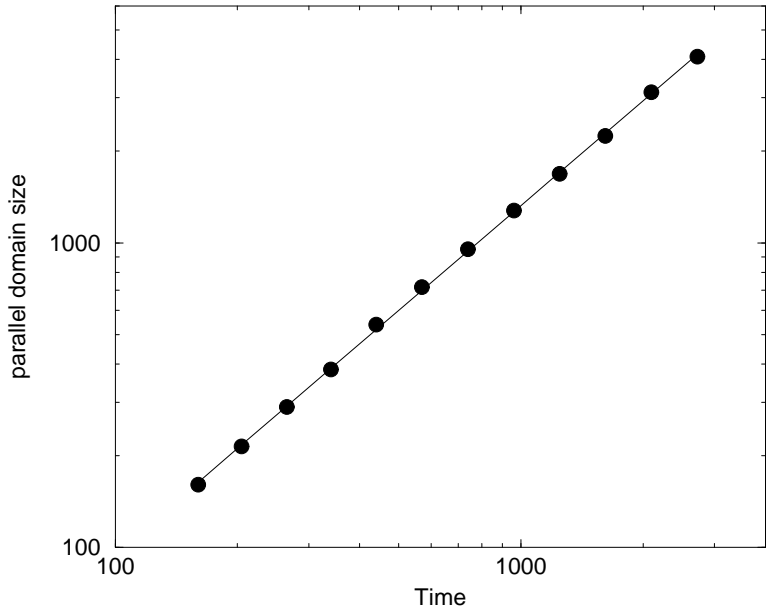}
\caption{Domain size $L_\parallel$ as a function of time, averaged over 10 
samples. The full line is a power fit, giving $L_\parallel(t)\sim t^{1.14}$. }
\end{center}
\end{figure}

Summarizing, we can say that, up to the simulational times we were able to 
reach, numerical data are largely compatible with our theoretical results.
In particular, the nontrivial leading behavior of $L_x(t)$, $L_y(t)$ and
$L_\parallel(t)$ are correctly reproduced, while the naive expectations
for the domain growth is sharply ruled out by the simulations.
Note that, of course, a longer simulational time would be desirable,
especially to check whether the curve of the energy develops a well
defined minimum and eventually starts increasing as $(\ln t)^{1/4}$. 
Unfortunately, as we have 
seen, the high values of $L_x$ and $L_\parallel$ make this impossible with
the system sizes we were able to reach, otherwise finite size effects 
would heavily come into play. Also for this reason,
in the next Section we propose a real experimental test of our 
analytical results.

\section{An experimental test of the two-dimensional results}

Theoretical and numerical results on nonconserved two-dimensional 
coarsening dynamics can be experimentally tested by means of thin films 
of uniaxial twisted nematic liquid crystal (TNLC) subjected to rapid 
thermal quenches. Since the classic experiments of Orihara, Ishibashi 
and Nagaya \cite{Orihara}, showing that a dynamical scaling compatible 
with the law $L(t)\sim t^{1/2}$ actually takes place in this system 
\cite{correction}, many other workers have successfully tested numerical 
and theoretical results on nonconserved coarsening in TNLC \cite{Yurke}.
In particular, let us note that this kind of system seems to be
particularly suitable for testing our analytic
calculation: indeed, it has been shown in \cite{Orihara} that the
scaling function describing the two-dimensional coarsening dynamics 
in TNLC is very well approximated by the analytic expression given by 
the OJK theory \cite{OJK}. Moreover, it has been explicitly checked 
\cite{Orihara} that the Allen-Cahn equation (\ref{allen}), describing 
the motion of an interface due to its curvature, holds to a 
very good degree of accuracy for TNLC. Our aim is to describe in 
this Section the basic experimental setup for TNLC and to propose 
a shear experiment on such systems, in order to test our nonstandard 
two-dimensional results in the case of simple, time-independent shear.

A typical TNLC cell is obtained by confining the sample of nematic liquid
crystal between two glass plates, previously prepared by rubbing them in 
two mutually perpendicular directions. In this way the orientations of the 
crystal molecules belonging to the two layers close to the plates have a  
relative rotation of $\pi/2$. At high temperature, in the isotropic phase, 
the boundary conditions only affect the system close to the boundaries, 
but when the crystal is quenched below the transition temperature (also 
called the {\it clearing point}), deep into the nematic phase, the 
alignment of the molecules with the boundary conditions on the plates 
extends into the bulk. In this way two different {\it states} appear, 
corresponding to the possibility of the molecules to rotate between the  
directions imposed by the two boundary plates in either a clockwise or an 
anti-clockwise sense. In other words, after the quench the TNLC cell 
develops  two equivalent states, which we may call left-handed and 
right-handed. Domains of the two states are separated by 
{\it disclination lines} \cite{Orihara}, defined as the points where 
the sense of rotation changes sign. The system is effectively 
two-dimensional and the dynamics of the left and right-handed domains 
is very well described by nonconserved coarsening dynamics. 

In order to reproduce the situation studied in the present paper, it
is necessary to shear the TNLC cell in a such a way that the shear
direction is {\it parallel} to the two plates (the flow direction is
of course parallel to them). Namely, the mutual orientation of the two
plates must not be changed in the experiment, while the orthogonal
walls must be moved in order to create the shear. In this way our
$(xy)$ plane would be parallel to the rubbed glass plates.

Given that a vital condition for testing our asymptotic results,  
in the case of time-independent shear, is the
possibility to shear the system for a long time, it seem to us that a
linear geometry is probably unsuitable for such an experiment. 
On the contrary, a circular setup may be more convenient: by taking
two circular glass plates, rubbed tangentially and radially, 
it is possible to create a cell whose wall, orthogonal
to the plates, can now be rotated indefinitely. In order to create 
the shear it is necessary to place a fixed cylinder at the center of the
system. In this way the material in contact with
this cylinder is stationary, while the layers close to the outer walls move
with a given tangential velocity $u_0$, creating a velocity profile given by
\beq
u(r) = \frac{u_0 R_0}{R_0^2-R_c^2} \, \left( r - \frac{R_c^2}{r} \right) 
\ \ \ , \ \ \ R_c<r<R_0 \ ,
\eeq
where $R_0$ and $R_c$ are the radius of the cell and of the internal cylinder,
respectively. If $R_c - R_0 \ll R_0$, it is possible to produce
a flow identical to the one studied in the present work and to study the 
long time dynamics of the domains under shear. Indeed, by setting 
$r=R_c +y$, we have
\beq
u(y)= \frac{2 u_0 R_0}{R_0^2-R_c^2} \  y \ \ \ , \ \ \ y\ll R_c  \ ,
\eeq
to be compared with relation (\ref{flow}).

Finally, testing our results in the case of oscillatory shear
should be easier from the experimental point of view, since the
periodicity of the shear function allows for the simpler linear 
geometry. As we have seen, the main growth follows a $t^{1/2}$ law,
modulated by some oscillations in the longitudinal direction.
In particular, it should not be difficult to test whether the 
ratio of perpendicular and parallel domain sizes satisfies the 
relation
\beq
\frac{L_\perp}{L_\parallel} \sim \frac{\omega}{\gamma} \ ,
\eeq
in the regime where $\gamma \gg \omega$.

\section{Conclusions}

In this paper we have analytically studied the effect of a shear flow
on phase ordering, for a statistical system with nonconserved scalar 
order parameter. We have developed a self-consistent anisotropic 
version of the OJK 
approximation, by means of which we have calculated the growth exponents
for time-independent shear in two and 
three dimensions (relations (\ref{end2d}) and (\ref{end3d})), 
and we have found the scaling form of the equal-time two-point correlation
function in both cases (relations (\ref{sf}), (\ref{assi}) and (\ref{calmo})).
While for $d=3$ our results are consistent with some simple scaling
arguments and with the results obtained for conserved dynamics in 
the limit of large dimension $N$ of the order parameter, 
in $d=2$ we find that domain growth is so heavily affected by the
shear, that the domains experience a narrowing which in principle makes 
their thickness vanish in the limit $t\to\infty$. However, as we have 
pointed out, our calculation is likely to break down for very long times, 
when the interface and the domain thickness are of the same order. 
What happens beyond this stage is still unclear to us: it is possible 
that a time-dependent steady state develops, with very narrow domains 
coalescing and giving rise to new thicker domains, which start 
narrowing again. Another possible scenario is that when 
$L_\bot\sim \xi$ domains start breaking and stretching again, giving 
rise to a steady state like the one depicted in \cite{Ohta}. Further work 
is needed to clarify this point and it is to be hoped that experiments 
on twisted nematic liquid crystals, as described in the last Section, 
will lead to a deeper understanding of this problem.

We have also studied the case of an oscillatory shear in two
dimensions, finding a standard $t^{1/2}$ growth, modulated by 
periodic oscillations which occur in opposition of phase
for the parallel and perpendicular direction.
Interestingly enough, all our results in this case are largely
independent on the particular form of the shear rate oscillations.

It is important to note that, in two dimensions, our results
satisfy the topological constraint on the growth of the scale
area, both in the time-independent and oscillatory case. This
fact, together with the results of our numerical simulations,
strongly support the validity of our method in the study of
coarsening systems under shear.

Of course, it would be very interesting to know  whether some of
our results (in particular in dimension two) are preserved for
conserved dynamics, which is the relevant case for describing spinodal 
decomposition in binary fluids. Unfortunately, the OJK approximation
cannot be used in this case, since the very starting point, the
Allen-Cahn equation for the interface motion, does not hold when the
order parameter is conserved. It is therefore still unclear how to
go beyond the large-$N$ limit in the context of spinodal decomposition 
under shear.

\acknowledgements
It is a pleasure to thank F. Colaiori, I. Giardina and F. Thalmann for useful 
discussions. This work was supported by EPSRC under grant GR/L97698 
(AC and AB), and by Funda\c c\~ao para a Ci\^encia e a Tecnologia (RT).


\begin{references}
\bibitem{correction} The actual value of the dynamical exponent found
by Orihara and Ishibashi was $1/z=0.44$.
\end{references}
\end{document}